# The Evolution of Galaxies by the Incompatibility between Dark Matter and Baryonic Matter*

In this paper, the evolution of galaxies is by the incompatibility between dark matter and baryonic matter. Due to the structural difference, baryonic matter and dark matter are incompatible to each other as oil droplet and water in emulsion. In the interfacial zone between dark matter and baryonic matter, this incompatibility generates the modification of Newtonian dynamics to keep dark matter and baryonic matter apart. The five periods of baryonic structure development in the order of increasing incompatibility are the free baryonic matter, the baryonic droplet, the galaxy, the cluster, and the supercluster periods. The transition to the baryonic droplet generates density perturbation in the CMB. In the galaxy period, the first-generation galaxies include elliptical, normal spiral, barred spiral, irregular, and dwarf spheroidal galaxies. In the cluster period, the second-generation galaxies include modified giant ellipticals, cD, evolved S0, dwarf elliptical, BCD, and tidal dwarf galaxies. The whole observable expanding universe behaves as one unit of emulsion with increasing incompatibility between dark matter and baryonic matter. The properties of dark matter and baryonic matter are based on cosmology derived from the two physical structures: the space structure and the object structure. Baryonic matter can be described by the periodic table of elementary particles.

Contents

## Introduction

Dark matter has been detected only indirectly by means of its gravitational effects astronomically. Dark matter as weakly interacting massive particles (WIMPs) has not been detected directly on the earth. This paper proposes that the absence of the direct detection of dark matter on the earth is due to the incompatibility between baryonic matter and dark matter, analogous to incompatible oil and water. The incompatibility means that when baryonic matter and dark matter are far away from each other, they attract each other through gravity, but when they are close together, they repel each other. The incompatibility leads to distribution of the dark matter and baryonic matter in different regions like emulsion. Most dark matter fermions form dark matter halos outside of baryonic galaxies. Some dark matter fermions remain inside of galaxies, such as the Milky Way Galaxy. The dark matter particles in the Milky Way Galaxy form the thin dark matter halos around star systems, such as the solar system. The repulsion in the thin dark matter halo causes Pioneer 10 and 11 spacecrafts in the outer space of the solar system to have anomalous acceleration toward the sun.

The properties of dark matter and baryonic matter are based on cosmology derived from the two physical structures: the space structure and the object structure. Baryonic matter can be described by the periodic table of elementary particles. Section 1 describes the two physical structures. Section 2 describes cosmology. The periodic table of elementary particles is described in Section 3. The evolution of galaxies is stated in Section 4.

# 1. The Two Physical Structures

## Introduction

The unified theory of physics unifies various phenomena in our observable universe and other universes. The unified theory of physics is derived from the unified universe [1] [2]. In the unified universe, all universes are governed by the cosmic code of the unified universe. Different universes are the different expressions of the same cosmic code of the unified universe. The cosmic code consists of two parts as the two physical structures: the space structure and the object structure.

### 1.1. The Space Structure

The first part of the physical structures is the space structure. The space structure [3] [4] consists of attachment space (denoted as 1) and detachment space (denoted as 0). Attachment space attaches to object permanently with zero speed or reversibly at the speed of light. Detachment space irreversibly detaches from the object at the speed of light. Attachment space relates to rest mass, while detachment space relates to kinetic energy. Different stages of our universe have different space structures.

The cosmic origin of detachment space is the cosmic radiation from the particle-antiparticle annihilation that initiates the inflation as shown later. Some objects in 4D-attachment space, denoted as $1_4$, convert into the cosmic radiation in 4D-detachment space, denoted as $0_4$. Cosmic radiation cannot permanently attach to a space.

$$\textit{some objects in } 1_4 \longrightarrow \textit{the cosmic radiation in } 0_4 \qquad (1)$$

The combination of attachment space (1) and detachment space (0) brings about three different space structures: miscible space, binary partition space, and binary lattice space for four-dimensional space-time as below.

$$(1)_n \textit{ attachment space } + (0)_n \det\textit{achment space } \xrightarrow{combination}$$

$$(1\ 0)_n \textit{ binary lattice space}, (1+0)_n \textit{ miscible space, or } (1)_n(0)_n \textit{ binary partition space} \qquad (2)$$

Binary lattice space, $(1\ 0)_n$, consists of repetitive units of alternative attachment space and detachment space. Thus, binary lattice space consists of multiple quantized units of attachment space separated from one another by detachment space. In miscible space, attachment space is miscible to detachment space, and there is no separation of attachment space and detachment space. Binary partition space, $(1)_n(0)_n$, consists of separated continuous phases of attachment space and detachment space.

Binary lattice space consists of multiple quantized units of attachment space separated from one another by detachment space. An object exists in multiple quantum states separated from one another by detachment space. Binary lattice space is the space for wavefunction. In wavefunction,

$$|\Psi\rangle = \sum_{i=1}^{n} c_i |\phi_i\rangle \quad , \qquad (3)$$

Each individual basis element, $|\phi_i\rangle$, attaches to attachment space, and separates from the adjacent basis element by detachment space. Detachment space detaches from object. Binary lattice space with n units of four-dimensional, $(0\ 1)_n$, contains n units of basis elements.

Neither attachment space nor detachment space is zero in binary lattice space. The measurement in the uncertainty principle in quantum mechanics is essentially the measurement of attachment space and momentum in binary lattice space: large momentum has small non-zero attachment space, while large attachment space has low non-zero momentum. In binary lattice space, an entity is both in constant motions as wave for detachment space and in stationary state as a particle for attachment space, resulting in the wave-particle duality.

Detachment space contains no object that carries information. Without information, detachment space is outside of the realm of causality. Without causality, distance (space) and time do not matter to detachment space, resulting in non-localizable and non-countable space-time. The requirement for the system (binary lattice space) containing non-localizable and non-countable detachment space is the absence of net information by any change in the space-time of detachment space. All changes have to be coordinated to result in zero net information. This coordinated non-localized binary lattice space corresponds to nilpotent space. All changes in energy, momentum, mass, time, space have to result in zero as defined by the generalized nilpotent Dirac equation by B. M. Diaz and P. Rowlands [5].

$$(\mp \mathbf{k}\partial/\partial t \pm \mathbf{i}\nabla + \mathbf{j}m)(\pm i\mathbf{k}E \pm \mathbf{i}\mathbf{p} + \mathbf{j}m)\exp i(-Et + \mathbf{p}.\mathbf{r}) = 0 \quad , \tag{4}$$

where E, **p**, m, t and **r** are respectively energy, momentum, mass, time, space and the symbols ± 1, ± *i*, ± *i*, ± *j*, ± *k*, ± **i**, ± **j**, ± **k**, are used to represent the respective units required by the scalar, pseudoscalar, quaternion and multivariate vector groups. The changes involve the sequential iterative path from nothing (nilpotent) through conjugation, complexification, and dimensionalization. The non-local property of binary lattice space for wavefunction provides the violation of Bell inequalities [6] in quantum mechanics in terms of faster-than-light influence and indefinite property before measurement. The non-locality in Bell inequalities does not result in net new information.

In binary lattice space, for every detachment space, there is its corresponding adjacent attachment space. Thus, no part of the object can be irreversibly separated from binary lattice space, and no part of a different object can be incorporated in binary lattice space. Binary lattice space represents coherence as wavefunction. Binary lattice space is for coherent system. Any destruction of the coherence by the addition of a different object to the object causes the collapse of binary lattice space into miscible space. The collapse is a phase transition from binary lattice space to miscible space.

$$((0)(1))_n \xrightarrow{collapse} (0+1)_n \tag{5}$$

*binary lattice space*      *miscible space*

Another way to convert binary lattice space into miscible space is gravity. Penrose [7] pointed out that the gravity of a small object is not strong enough to pull different states into one location. On the other hand, the gravity of large object pulls different quantum states into one location to become miscible space. Therefore, a small object without outside interference is always in binary lattice space, while a large object is never in binary lattice space.

The information in miscible space is contributed by the combination of both attachment space and detachment space, so information can no longer be non-localize. Any value in miscible space is definite. All observations in terms of measurements bring about the collapse of wavefunction, resulting in miscible space that leads to eigenvalue as definite quantized value. Such collapse corresponds to the appearance of eigenvalue, E, by a measurement operator, H, on a wavefunction, Ψ.

$$H\Psi = E\Psi \quad , \tag{6}$$

In miscible space, attachment space is miscible to detachment space, and there is no separation of attachment space and detachment space. In miscible space, attachment space contributes zero speed, while detachment space contributes the speed of light. A massless particle, such as photon, is on detachment space continuously, and detaches from its own space continuously. For a moving massive particle consisting of a rest massive part and a massless part, the massive part with rest mass, $m_0$, is in attachment space, and the massless part with kinetic energy, $K$, is in detachment space. The combination of the massive part in attachment space and massless part in detachment leads to the propagation speed in between zero and the speed of light.

To maintain the speed of light constant for a moving particle, the time (t) in moving particle has to be dilated, and the length (L) has to be contracted relative to the rest frame.

$$t = = t_0 \Big/ \sqrt{1 - v^2/c^2} \ = \ t_0 \gamma,$$
$$L = L_0 / \gamma, \qquad (7)$$
$$E = K + m_0 c^2 = \gamma m_0 c^2$$

where $\gamma = 1/\sqrt{1 - v^2/c^2}$ is the Lorentz factor for time dilation and length contraction, $E$ is the total energy and $K$ is the kinetic energy.

Binary partition space, $(1)_n(0)_n$, consists of separated continuous phases of attachment space and detachment space. It is for extreme force fields under extreme conditions such as near the absolute zero temperature or extremely high pressure. It will be discussed later to explain extreme phenomena such as superconductivity and black hole.

### 1.2. The Object Structure

The second part of the physical structures is the object structure. The object structure consists of 11D membrane ($3_{11}$), 10D string ($2_{10}$), variable D particle ($1_{4 \text{ to } 10}$), and empty object ($0_{4 \text{ to } 11}$). Different universes and different stages of a universe can have different expressions of the object structure. For an example, the four stages in the evolution of our universe are the 11D membrane universe (the strong universe), the dual 10D string universe (the gravitational pre-universe), the dual 10D particle universe (the charged pre-universe), and the dual 4D/variable D particle universe (the current universe).

The transformation among the objects is through the dimensional oscillation [2] that involves the oscillation between high dimensional space-time and low dimensional space-time. The vacuum energy of the multiverse background is about the Planck energy. Vacuum energy decreases with decreasing dimension number. The vacuum energy of 4D space-time is zero. With such vacuum energy differences, the local dimensional oscillation between high and low space-time dimensions results in local eternal expansion-contraction [8] [9] [10]. Eternal expansion-contraction is like harmonic oscillator, oscillating between the Planck vacuum energy and the lower vacuum energy.

For the dimensional oscillation, contraction occurs at the end of expansion. Each local region in the universe follows a particular path of the dimensional oscillation. Each path is marked by particular set of force fields. The path for our universe is marked by the strong force, gravity-antigravity, charged electromagnetism, and asymmetrical weak force, corresponding to the four stages of the cosmic evolution.

The vacuum energy differences among space-time dimensions are based on the varying speed of light. Varying speed of light has been proposed to explain the horizon problem of cosmology [11][12]. The proposal is that light traveled much faster in the distant past to allow distant regions of the expanding universe to interact since the beginning of the universe. Therefore, it was proposed as an alternative to cosmic inflation. J. D. Barrow [13] proposes that the time dependent speed of light varies as some power of the expansion scale factor $a$ in such way that

$$c(t) = c_0 a^n \qquad (8)$$

where $c_0 > 0$ and $n$ are constants. The increase of speed of light is continuous.

In this paper, varying dimension number (VDN) relates to quantized varying speed of light (QVSL), where the speed of light is invariant in a constant space-time dimension number, and the speed of light varies with varying space-time dimension number from 4 to 11.

$$c_D = c/\alpha^{D-4}, \qquad (9)$$

where $c$ is the observed speed of light in the 4D space-time, $c_D$ is the quantized varying speed of light in space-time dimension number, D, from 4 to 11, and $\alpha$ is the fine structure constant for electromagnetism. Each dimensional space-time has a specific speed of light. (Since from the beginning of our observable universe, the space-time dimension has always been four, there is no observable varying speed of light in our observable universe.) The speed of light increases with the increasing space-time dimension number D.

In special relativity, $E = M_0 c^2$ modified by Eq. (9) is expressed as

$$E = M_0 \cdot (c^2/\alpha^{2(D-4)}) \qquad (10a)$$
$$= (M_0/\alpha^{2(d-4)}) \cdot c^2. \qquad (10b)$$

Eq. (10a) means that a particle in the D dimensional space-time can have the superluminal speed $c/\alpha^{D-4}$, which is higher than the observed speed of light $c$, and has the rest mass $M_0$. Eq. (10b) means that the same particle in the 4D space-time with the observed speed of light acquires $M_0/\alpha^{2(d-4)}$ as the rest mass, where d = D. D in Eq. (10a) is the space-time dimension number defining the varying speed of light. In Eq. (10b), d from 4 to 11 is "mass dimension number" defining varying mass. For example, for D = 11, Eq. (10a) shows a superluminal particle in eleven-dimensional space-time, while Eq. (10b) shows that the speed of light of the same particle is the observed speed of light with the 4D space-time, and the mass dimension is eleven. In other words, 11D space-time can transform into 4D space-time with 11d mass dimension. 11D4d in Eq. (10a) becomes 4D11d in Eq. (10b) through QVSL. QVSL in terms of varying space-time dimension number, D, brings about varying mass in terms of varying mass dimension number, d.

The QVSL transformation transforms both space-time dimension number and mass dimension number. In the QVSL transformation, the decrease in the speed of light leads to the decrease in space-time dimension number and the increase of mass in terms of increasing mass dimension number from 4 to 11,

$$c_D = c_{D-n} / \alpha^{2n}, \tag{11a}$$

$$M_{0,D,d} = M_{0,D-n,\ d+n} \alpha^{2n}, \tag{11b}$$

$$D, d \xrightarrow{QVSL} (D \mp n),\ (d \pm n) \tag{11c}$$

where D is the space-time dimension number from 4 to 11 and d is the mass dimension number from 4 to 11. For example, in the QVSL transformation, a particle with 11D4d is transformed to a particle with 4D11d. In terms of rest mass, 11D space-time has 4d with the lowest rest mass, and 4D space-time has 11d with the highest rest mass.

Rest mass decreases with increasing space-time dimension number. The decrease in rest mass means the increase in vacuum energy, so vacuum energy increases with increasing space-time dimension number. The vacuum energy of 4D particle is zero, while 11D membrane has the Planck vacuum energy. Such vacuum energies are the alternatives for the Higgs bosons, which have not been found. The decrease in vacuum energy is equivalent to the absorption of the Higgs boson, while the increase in vacuum energy is equivalent to the emission of the Higgs boson.

Since the speed of light for > 4D particle is greater than the speed of light for 4D particle, the observation of > 4D particles by 4D particles violates casualty. Thus, > 4D particles are hidden particles with respect to 4D particles. Particles with different space-time dimensions are transparent and oblivious to one another, and separate from one another if possible.

### 1.3. Summary

The unified theory of physics is derived from the two physical structures. Different universes in different developmental stages are the different expressions of the same two physical structures. The two physical structures are the space structure and the object structure. Relating to rest mass, attachment space attaches to object permanently with zero speed or reversibly at the speed of light. Relating to kinetic energy, detachment space irreversibly detaches from the object at the speed of light. In our observable universe, the space structure consists of three different combinations of attachment space and detachment space, describing three different phenomena: quantum mechanics, special relativity, and the extreme force fields. The object structure consists of 11D membrane ($3_{11}$), 10D string ($2_{10}$), variable D particle ($1_{\leq 10}$), and empty object (0). The space structure includes attachment space (1) and detachment space (0). The transformation among the objects is through the dimensional oscillation that involves the oscillation between high dimensional space-time with high vacuum energy and low dimensional space-time with low vacuum energy. Our observable universe with 4D space-time has zero vacuum energy.

# 2. Cosmology

## Introduction

Before the current universe, the pre-universe is in the three different stages in chronological order: the strong pre-universe, the gravitational pre-universe, and the charged pre-universe. The strong pre-universe has only one force: the strong force. The gravitational pre-universe has two forces: the strong and the gravitational forces. The charged pre-universe has three forces: the strong, the gravitational, and the electromagnetic forces. All three forces in the pre-universes are in their primitive forms unlike the finished forms in our observable universe. The asymmetrical weak interaction comes from the formation of the current asymmetrical dual universe. Such 4-stage cosmology for our universe explains the origin of the four force fields in our observable universe.

### 2.1. The Strong Pre-Universe

| Dual universe | Object structure | Space structure | Force |
|---|---|---|---|
| no | 11D membrane | attachment space | pre-strong |

Many different universes can emerge from the multiverse background, which has the simplest and most primitive structure [1] [2]. As in Einstein's static universe, the time in the multiverse background has no beginning. Different parts of the background have potential to undergo local inhomogeneity to develop different universes with different object structures, space structures, and vacuum energies. The multiverse background is the strong pre-universe. It is the homogeneous static universe, consisting of 11D (space-time dimensional) positive energy membrane and negative energy anti-membrane, denoted as $3_{11}$ $3_{-11}$, as proposed by Mongan [14]. The only force among the membranes is the pre-strong force, s, as the predecessor of the strong force. It is from the quantized vibration of the membranes to generate the reversible process of the absorption-emission of the massless particles among the membranes. The pre-strong force mediates the reversible absorption-emission in the flat space. The pre-strong force is the same for all membranes, so it is not defined by positive or negative sign. It does not have gravity that causes instability and singularity [15], so the initial universe remains homogeneous, flat, and static. This initial universe provides the globally stable static background state for an inhomogeneous eternal universe in which local regions undergo expansion-contraction [15].

### 2.2. The Gravitational Pre-Universe

| Dual universe | Object structure | Space structure | Forces |
|---|---|---|---|
| dual | 10D string | attachment space | pre-strong, pre-gravity |

In certain regions of the 11D membrane universe, the local expansion takes place by the transformation from 11D-membrane into 10D-string. The expansion is the result of the vacuum energy difference between 11D membrane and 10D string. With the emergence of empty object ($0_{11}$), 11D membrane transforms into 10D string warped with virtue particle as pregravity.

$$3_{11} s + 0_{11} \longleftrightarrow 2_{10} s 1_1 = 2_{10} s g^+ \qquad (12)$$

where $3_{11}$ is the 11D membrane, s is the pre-strong force, $0_{11}$ is the 11D empty object, $2_{10}$ is 10D string, $1_1$ is one dimensional virtue particle as g, pre-gravity. Empty object corresponds to the anti-De Sitter bulk space in the Randall-Sundrum model [16]. In the same way, the surrounding object can extend into empty object by the decomposition of space dimension as described by Bounias and Krasnoholovets [17], equivalent to the Randall-Sundrum model. The g is in the bulk space, which is the warped space (transverse radial space) around $2_{10}$. As in the AdS/CFT duality [18] [19] [20], the pre-strong force has 10D dimension, one dimension lower than the 11D membrane, and is the conformal force defined on the conformal boundary of the bulk space. The pre-strong force mediates the reversible absorption-emission process of membrane (string) units in the flat space, while pregravity mediates the reversible condensation-decomposition process of mass-energy in the bulk space.

Through symmetry, antistrings form 10D antibranes with anti-pregravity as $2_{-10} g^-$, where $g^-$ is anti-pregravity.

$$3_{-11} s + 0_{-11} \longleftrightarrow 2_{-10} s \, 1_{-1} = 2_{-10} s g^- \qquad (13)$$

Pregravity can be attractive or repulsive to anti-pregravity. If it is attractive, the universe remains homogeneous. If it is repulsive, n units of $(2_{10})_n$ and n units of $(2_{-10})_n$ are separated from each other.

$$((s 2_{10}) g^+)_n (g^- (s 2_{-10}))_n \qquad (14)$$

The universe with pregravity and anti-pregravity is the dual 10D string universe, which leads to the evolution of our observable universe. The dual 10D string universe consists of two parallel universes with opposite energies: 10D strings with positive energy and 10D antistrings with negative energy. The two universes are separated by the bulk space, consisting of pregravity and anti-pregravity. Such dual universe separated by bulk space appears in the ekpyrotic universe model [21] [22].

### 2.3. The Charged Pre-Universe

| Dual universe | Object structure | Space structure | Forces |
|---|---|---|---|
| dual | 10D particle | attachment space | pre-strong, pre-gravity, pre-electromagnetic |

When the local expansion stops, through the dimensional oscillation, the contraction begins to force the dual 10D string universe to contract to the original state, resulting in the coalescence of the two universes. The coalescence allows the two universes to mix. The first path of such mixing is the string-antistring annihilation, resulting in disappearance of the dual universe and the return to the multiverse background. The outcome is the completion of one oscillating cycle.

The second path allows the continuation of the dual universe in another form without the mixing of positive energy and negative energy. Such dual universe is possible by the emergence of the pre-charge force, the predecessor of electromagnetism with positive and negative charges. The mixing becomes the mixing of positive charge and negative charge instead of positive energy and negative energy, resulting in the preservation of the dual universe with the positive energy and the negative energy. Our universe follows the second path as described below in details.

During the coalescence for the second path, the two universes coexist in the same space-time, which is predicted by the Santilli isodual theory [23]. Antiparticle for our positive energy universe is described by Santilli as follows, "this identity is at the foundation of the perception that antiparticles "appear" to exist in our space, while in reality they belong to a structurally different space coexisting within our own, thus setting the foundations of a "multidimensional universe" coexisting in the same space of our sensory perception" (Ref. 23, p. 94). Antiparticles in the positive energy universe actually come from the coexisting negative energy universe.

The mixing process follows the isodual hole theory that is the combination of the Santilli isodual theory and the Dirac hole theory. In the Dirac hole theory that is not symmetrical, the positive energy observable universe has an unobservable infinitive sea of negative energy. A hole in the unobservable infinitive sea of negative energy is the observable positive energy antiparticle.

In the dual 10D string universe, one universe has positive energy strings with pregravity, and one universe has negative energy antistrings with anti-pregravity. For the mixing of the two universes during the coalescence, a new force, the pre-charged force, emerges to provide the additional distinction between string and antistring. The pre-charged force is the predecessor of electromagnetism. Before the mixing, the positive energy string has positive pre-charge ($e^+$), while the negative energy antistring has negative pre-charge ($e^-$). During the mixing when two 10D string universes coexist, a half of positive energy strings in the positive energy universe move to the negative energy universe, and leave the Dirac holes in the positive energy universe. The negative energy antistrings that move to fill the holes become positive energy antistrings with negative pre-charge in the positive energy universe. In terms of the Dirac hole theory, the unobservable infinitive sea of negative energy is in the negative energy universe from the perspective of the positive energy universe before the mixing. The hole is due to the move of the negative energy antistring to the positive energy universe from the perspective of the positive energy universe during the mixing, resulting in the positive energy antistring with negative pre-charge in the positive energy universe.

In the same way, a half of negative energy antistrings in the negative energy universe moves to the positive energy universe, and leave the holes in the negative energy universe. The positive energy strings that move to fill the holes become negative energy strings with positive pre-charge in the negative energy universe. The result of the mixing is that both positive energy universe and the negative energy universe have strings-antistrings. The existence of the pre-charge provides the distinction between string and antistring in the string-antistring.

At that time, the space (detachment space) for radiation has not appeared in the universe, so the string-antistring annihilation does not result in radiation. The string-antistring annihilation results in the replacement of the string-antistring as the 10D string-antistring, ($2_{10}$ $2_{-10}$) by the 10D particle-antiparticle ($1_{10}$ $1_{-10}$). The 10D particles-antiparticles have the multiple dimensional Kaluza-Klein structure with variable space dimension number without the requirement for a fixed space dimension number for string-antistring. After the mixing, the dual 10D particle-antiparticle universe separated by pregravity and anti-pregravity appears as below.

$$((s1_{10}\ e^+\ e^-\ 1_{-10}\ s)g^+)_n \quad (g^-(s1_{10}\ e^+\ e^-\ 1_{-10}\ s))_n , \quad (15)$$

where s and e are the pre-strong force and the pre-charged force in the flat space, g is pregravity in the bulk space, and $1_{10}$ $1_{-10}$ is the particle-antiparticle. The dual 10D particle universe has particles, while the multiverse background (11D- membrane universe) has membranes, so the multiverse background and the dual 10D particle universe are completely transparent and oblivious to each other. The result is the free charged dual 10D particle-antiparticle universe.

The dual 10D particle universe consists of two parallel particle-antiparticle universes with opposite energies and the bulk space separating the two universes. There are four space regions: the positive energy particle-antiparticle space region, the pregravity bulk space region, the negative energy particle-antiparticle space region, and the anti-pregravity bulk space region.

**2.4. The Current Universe**

|  | Object structure | Space structure | Forces |
| --- | --- | --- | --- |
| The light universe | 4D particle | attachment space and detachment space | strong, gravity, electromagnetic, and weak |
| The dark universe | variable D between 4 and 10 particle | attachment space | pre-strong, gravity, pre-electromagnetic |

The formation of our current universe follows immediately after the formation of the charged pre-universe through the asymmetrical dimensional oscillations, leading to the asymmetrical dual universe consisting of the light universe with kinetic energy and light and the dark universe without kinetic energy and light. Our observable universe is the light universe, whose formation involves the immediate transformation from 10D to 4D, resulting in the inflation as shown later. The formation of the dark universe involves the slow dimensional oscillation between 10D and 4D. The asymmetrical dual universe is manifested as the asymmetry in the weak interaction in our observable universe as follows.

$$((s1_4\ e^+w^+\ e^-w^-\ 1_{-4}\ s)g^+)_n \quad (g^-(s1_{\leq 10}\ e^+w^+\ e^-w^-\ 1_{\geq -10}\ s))_n \quad (16)$$

where s, g, e, and w are the strong force, gravity, electromagnetism, and weak interaction, respectively for the observable universe, and where $1_4 1_{-4}$ and $1_{\leq 10} 1_{\geq -10}$ are 4D particle-antiparticle for the light universe and variable D particle-antiparticle for the dark universe, respectively.

In summary, the whole process of the local dimensional oscillations leading to our observable universe is illustrated as follows.

$$membrane\ universe \xleftrightarrow{between\ 11D\ and\ 10D} dual\ string\ universe \xleftrightarrow{coalescence,\ annihilation}$$
$$3_{11}\, s\, s\, 3_{-11} \qquad\qquad ((s2_{10})g^+)_n (g^-(s2_{-10}))_n$$

$$dual\ 10D\ particle\ universe \xleftrightarrow{between\ 10D\ and\ 4D} dual\ 4D/variable\ D\ particle\ universe$$
$$((s1_{10}\,e^+\,e^-\,1_{-10}\,s)g^+)_n\,(g^-(s1_{10}\,e^+\,e^-\,1_{-10}\,s))_n \qquad ((s1_4\,e^+w^+e^-w^-\,1_{-4}\,s)g^+)_n\,(g^-(s1_{\leq 10}\,e^+w^+e^-w^-\,1_{\geq -10}\,s))_n$$

where s, e, and w are in the flat space, and g is in the bulk space. Each stage generates one force, so the four stages produce the four different forces: the strong force, gravity, electromagnetism, and the weak interaction, sequentially. Gravity appears in the first dimensional oscillation between the 11 dimensional membrane and the 10 dimensional string. The asymmetrical weak force appears in the asymmetrical second dimensional oscillation between the ten dimensional particle and the four dimensional particle. Charged electromagnetism appears as the force in the transition between the first and the second dimensional oscillations. The cosmology explains the origins of the four forces. To prevent the charged pre-universe to reverse back to the previous pre-universe, the charge pre-universe and the current universe overlap to a certain degree as shown in the overlapping between the electromagnetic interaction and the weak interaction to form the electroweak interaction.

| Four-Stage Universe | Universe | Object Structure | Space Structure | Force |
|---|---|---|---|---|
| **Strong Pre-Universe** | single | 11D membrane | attachment space | pre-strong |
| **Gravitational Pre-Universe** | dual | 10D string | attachment space | pre-strong, pre-gravity |
| **Charged Pre-Universe** | dual | 10D particle | attachment space | pre-strong, pre-gravity, pre-electromagnetic |
| **Current Universe** | dual | | | |
| light universe | | 4D particle | attachment space and detachment space | strong, gravity, electromagnetic, and weak |
| dark universe | | variable D between 4 and 10 particle | attachment space | pre-strong, gravity, pre-electromagnetic |

The formation of the dark universe involves the slow dimensional oscillation between 10D and 4D. The dimensional oscillation for the formation of the dark universe involves the stepwise two-step transformation: the QVSL transformation and the varying supersymmetry transformation. In the normal supersymmetry transformation, the repeated application of the fermion-boson transformation carries over a boson (or fermion) from one point to the same boson (or fermion) at another point at the same mass. In the "varying supersymmetry transformation", the repeated application of the fermion-boson transformation carries over a boson from one point to the boson at another point at different mass dimension number in the same space-time number. The repeated varying supersymmetry transformation carries over a boson $B_d$ into a fermion $F_d$ and a fermion $F_d$ to a boson $B_{d-1}$, which can be expressed as follows

$$M_{d,F} = M_{d,B}\, \alpha_{d,B}, \tag{17a}$$

$$M_{d-1,B} = M_{d,F}\, \alpha_{d,F}, \tag{17b}$$

where $M_{d,B}$ and $M_{d,F}$ are the masses for a boson and a fermion, respectively, d is the mass dimension number, and $\alpha_{d,B}$ or $\alpha_{d,F}$ is the fine structure constant that is the ratio between the masses of a boson and its fermionic partner. Assuming $\alpha_{d,B}$ or $\alpha_{d,F}$, the relation between the bosons in the adjacent dimensions or n dimensions apart (assuming α's are the same) then can be expressed as

$$M_{d,B} = M_{d+1,B}\, \alpha_{d+n}^{2}. \tag{17c}$$

$$M_{d,B} = M_{d+n,B}\, \alpha_{d+n}^{2n}. \tag{17d}$$

Eq. (18) show that it is possible to describe mass dimensions > 4 in the following way

$$F_5\ B_5\ F_6\ B_6\ F_7\ B_7\ F_8\ B_8\ F_9\ B_9\ F_{10}\ B_{10}\ F_{11}\ B_{11}, \tag{18}$$

where the energy of $B_{11}$ is the Planck energy. Each mass dimension between 4d and 11d consists of a boson and a fermion. Eq. (19) show a stepwise transformation that converts a particle with d mass dimension to d ± 1 mass dimension.

$$D, d \xleftarrow{\text{stepwise varying supersymmetry}} D, (d \pm 1) \tag{19}$$

The transformation from a higher mass dimensional particle to the adjacent lower mass dimensional particle is the fractionalization of the higher dimensional particle to the many lower dimensional particle in such way that the number of lower dimensional particles becomes $n_{d-1} = n_d / \alpha^2$. The transformation from lower dimensional particles to higher dimensional particle is a condensation. Both the fractionalization and the condensation are stepwise. For example, a particle with 4D (space-time) 10d (mass dimension) can transform stepwise into 4D9d particles. Since the supersymmetry transformation involves translation, this stepwise varying supersymmetry transformation leads to a translational fractionalization and translational condensation, resulting in expansion and contraction.

For the formation of the dark universe from the charged pre-universe, the negative energy universe has the 10D4d particles, which is converted eventually into 4D4d stepwise and slowly. It involves the stepwise two-step varying transformation: first the QVSL transformation, and then, the varying supersymmetry transformation as follows.

$$\text{stepwise two-step varying transformation}$$
$$(1) \quad D, d \xleftrightarrow{QVSL} (D \mp 1), (d \pm 1) \tag{20}$$
$$(2) \quad D, d \xleftrightarrow{varying\ supersymmetry} D, (d \pm 1)$$

The repetitive stepwise two-step transformations from 10D4d to 4D4d are as follows.

*The Hidden Dark Universe and the Observable Dark Universe with Dark Energy*

$$10D4d \to 9D5d \to 9D4d \to 8D5d \to 8D4d \to 7D5d \to \bullet\bullet\bullet\bullet \to 5D4d \to 4D5d \to 4D4d$$
$$\mapsto \quad the \quad hidden \quad dark \quad universe \leftarrow\mapsto dark\ energy \leftarrow$$

The dark universe consists of two periods: the hidden dark universe and the dark energy universe. The hidden dark universe composes of the > 4D particles. As mentioned before, particles with different space-time dimensions are transparent and oblivious to one another, and separate from one another if possible. Thus, > 4D particles are hidden and separated particles with respect to 4D particles in the light universe (our observable universe). The universe with > 4D particles is the hidden dark universe. The 4D particles transformed from hidden > 4D particles in the dark universe are observable dark energy for the light universe, resulting in the accelerated expanding universe. The accelerated expanding universe consists of the positive energy 4D particles-antiparticles and dark energy that includes the negative energy 4D particles-antiparticles and the antigravity. Since the dark universe does not have detachment space, the presence of dark energy is not different from the presence of the non-zero vacuum energy. In terms of quintessence, such dark energy can be considered the tracking quintessence [24] from the dark universe with the space-time dimension as the tracker. The tracking quintessence consists of the hidden quintessence and the observable quintessence. The hidden quintessence is from the hidden > 4D dark universe. The observable quintessence is from the observable 4D dark universe with 4D space-time.

For the formation of the light universe, the dimensional oscillation for the positive energy universe transforms 10D to 4D immediately. It involves the leaping two-step varying transformation, resulting in the light universe with kinetic energy. The first step is the space-time dimensional oscillation through QVSL. The second step is the mass dimensional oscillation through slicing-fusion.

$$\textit{leaping two-step varying transformation}$$

$$(1) \quad D, d \xleftrightarrow{QVSL} (D \mp n), (d \pm n) \tag{21}$$

$$(2) \quad D, d \xleftrightarrow{\textit{slicing-fusion}} D, (d \pm n) + (11 - d + n) DO's$$

The Light Universe

$$10D4d \xrightarrow{\textit{quick QVSL transformation}} 4D10d \xrightarrow{\textit{slicing with detachment space, inflation}}$$
$$\textit{dark matter } (4D10d + 4D9d + 4D8d + 4D7d + 4D6d + 4D5d) + \textit{baryonic matter } (4D4d) + \textit{cosmic radiation}$$
$$\rightarrow \textit{thermal cosmic expansion (the big bang)}$$

In the charged pre-universe, the positive energy universe has 10D4d, which is transformed into 4D10d in the first step through the QVSL transformation. The second step of the leaping varying transformation involves the slicing-fusion of particle. Bounias and Krasnoholovets [25] propose another explanation of the reduction of > 4 D space-time into 4D space-time by slicing > 4D space-time into infinitely many 4D quantized units surrounding the 4D core particle. Such slicing of > 4D space-time is like slicing 3-space D object into 2-space D object in the way stated by Michel Bounias as follows: "You cannot put a pot into a sheet without changing the shape of the 2-D sheet into a 3-D dimensional packet. Only a 2-D slice of the pot could be a part of sheet".

The slicing is by detachment space, as a part of the space structure, which consists of attachment space (denoted as 1) and detachment space (denoted as 0) as described earlier. Attachment space attaches to object permanently with zero speed or reversibly at the speed of light. Detachment space irreversibly detaches from the object at the speed of light. Attachment space relates to rest mass, while detachment space relates to kinetic energy. The cosmic origin of detachment space is the cosmic radiation from the particle-antiparticle annihilation that initiates the transformation. The cosmic radiation cannot permanently attach to a space.

The slicing of dimensions is the slicing of mass dimensions. 4D10d particle is sliced into seven particles: 4D10d, 4D9d, 4D8d, 4D7d, 4D6d, 4D5d, and 4D4d equally by mass. Baryonic matter is 4D4d, while dark matter consists of the other six types of particles (4D10d, 4D9d, 4D8d, 4D7d, 4D6d, and 4D5d) as described later. The mass ratio of dark matter to baryonic matter is 6 to 1 in agreement with the observation [26] showing the universe consists of 23% dark matter, 4% baryonic matter, and 73% dark energy.

Detachment space (0) involves in the slicing of mass dimensions. Attachment space is denoted as 1. For example, the slicing of 4D10d particles into 4D4d particles is as follows.

$$\left(1_{4+6}\right)_i \xrightarrow{\textit{slicing}} \left(1_4\right)_i + \sum_1^6 \left(\left(0_4\right)\left(1_4\right)\right)_{j,6} \tag{22}$$

$$> 4d \textit{ attachment space} \qquad 4d \textit{ core attachment space} \qquad 6 \textit{ types of } 4d \textit{ units}$$

The two products of the slicing are the 4d-core attachment space and 6 types of 4d quantized units. The 4d core attachment space surrounded by 6 types of many (j) 4D4d quantized units corresponds to the core particle surrounded by 6 types of many small 4d particles.

Therefore, the transformation from d to d – *n* involves the slicing of a particle with d mass dimension into two parts: the core particle with d – *n* dimension and the *n* dimensions that are separable from the core particle. Such *n* dimensions are denoted as *n* "dimensional orbitals", which become gauge force fields as described later. The sum of the number of mass dimensions for a particle and the number of dimensional orbitals (DO's) is equal to 11 (including gravity) for all particles with mass dimensions. Therefore,

$$F_d = F_{d-n} + (11 - d + n)\, \text{DO's} \qquad (23)$$

where $11 - d + n$ is the number of dimensional orbitals (DO's) for $F_{d-n}$. Thus, 4D10d particles can transformed into 4D10d, 4D9d, 4D8d, 4D7d, 4D6d, 4D5d, and 4D4d core particles, which have 1, 2, 3, 4, 5, 6, and 7 separable dimensional orbitals, respectively. Dark matter particle, 4D10d, has only gravity, while baryonic matter particle, 4D4d, has gravity and six other dimensional orbitals as gauge force fields as below.

The six > 4d mass dimensions (dimensional orbitals) for the gauge force fields and the one mass dimension for gravity are as in Figure 1.

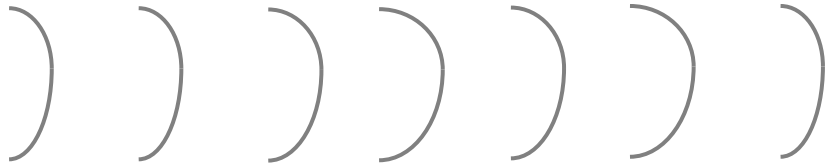

Figure 1. The seven force fields as > 4d mass dimensions (dimensional orbitals).

The dimensional orbitals of baryonic matter provide the base for the periodic table of elementary particles to calculate accurately the masses of all 4D elementary particles, including quarks, leptons, and gauge bosons as described later.

The lowest dimensional orbital is for electromagnetism. Baryonic matter is the only one with the lowest dimensional orbital for electromagnetism. With higher dimensional orbitals, dark matter does not have this lowest dimensional orbital. Without electromagnetism, dark matter cannot emit light, and is incompatible to baryonic matter, like the incompatibility between oil and water. The incompatibility between dark matter and baryonic matter leads to the inhomogeneity (like emulsion), resulting in the formation of galaxies, clusters, and superclusters as described later. Dark matter has not been found by direct detection because of the incompatibility.

In the light universe, the inflation is the leaping varying transformation that is the two-step inflation. The first step is to increase the rest mass as potential from higher space-time dimension to lower space-time dimension as expressed by Eq. (24a) from Eq. (11b).

$$D, d \xrightarrow{QVSL} (D \mp n), (d \pm n)$$
$$V_{D,d} = V_{D-n, d+n} \alpha^{2n}$$
$$\varphi = \text{collective } n's$$
$$V(\varphi) = V_{4D10d} \alpha^{-2\varphi}, \text{ where } \varphi \leq 0 \text{ from } -6 \text{ to } 0$$

(24a)

where $\alpha$ is the fine structure constant for electromagnetism. The ratio of the potential energies of 4D10d to that of 10D4d is $1/\alpha^{12}$. $\varphi$ is the scalar field for QVSL, and is equal to collective n's as the changes in space-time dimension number for many particles. The increase in the change of space-time dimensions from 4D decreases the potential as the rest mass. The region for QVSL is $\varphi \leq 0$ from -6 to 0. The QVSL region is for the conversion of the vacuum energy into the rest mass as the potential. The conversion of vacuum energy into potential is equivalent to the absorption of the Higgs boson, while the conversion of potential into vacuum energy is equivalent to the emission of the Higgs boson.

The second step is the slicing that occurs simultaneously with the appearance of detachment space that is the space for cosmic radiation (photon). Potential energy as massive 4D10d particles is converted into kinetic energy as cosmic radiation and massive matter particles (from 10d to 4d). It relates to the ratio between photon and matter in terms of the CP asymmetry between particle and antiparticle. The slight excess particle over antiparticle results in matter particle. The equation for the potential (V) and the scalar field ($\phi$) is as Eq. (24b) from Eq. (35) that expresses the ratio between photon and matter.

$$D, d \xrightarrow{\text{slicing}} D, (d-n)$$
$$V(\phi) = V_{4D10d} \alpha^{2\phi}, \text{ where } \phi \geq 0 \text{ from } 0 \text{ to } 2$$

(24b)

The ratio is $\alpha^4$, according to Eq. (35). The region for the slicing is $\phi \geq 0$ from 0 to 2. The slicing region is for the conversion of the potential energy into the kinetic energy.

The combination of Eq. (24a) and Eq. (24b) is as Eq. (24c).

$$V(\varphi, \phi) = V_{4D10d} (\alpha^{-2\varphi} + \alpha^{2\phi}),$$
$$\text{where } \varphi \leq 0 \text{ and } \phi \geq 0$$

(24c)

The graph for the two-step inflation is as Figure 2.

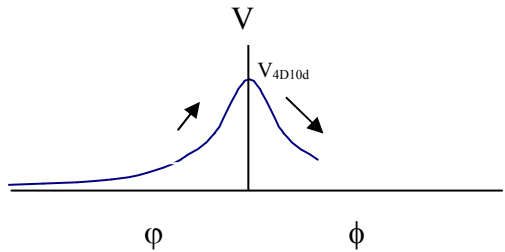

Figure 2. the two-step inflation

At the transition ($V_{4D10d}$) between the first step (QVSL) and the second step (slicing), the scalar field reverses its sign and direction. In the first step, the universe inflates by the decrease in vacuum energy. In the second step, the potential energy is converted into kinetic energy as cosmic radiation. The resulting kinetic energy starts the big bang, resulting in the expanding universe.

Toward the end of the cosmic contraction after the big crunch, the deflation occurs as the opposite of the inflation. The kinetic energy from cosmic radiation decreases, as the fusion occurs to eliminate detachment space, resulting in the increase of potential energy. At the end of the fusion, the force fields except gravity disappear, 4D10d particles appear, and then the scalar field reverses its sign and direction. The vacuum energy increases as the potential as the rest mass decreases for the appearance of 10D4d particles, resulting in the end of a dimensional oscillation as Figure 3 for the two-step deflation.

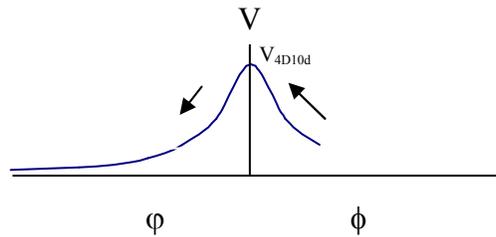

Figure 3. the two-step deflation

The end of the two-step deflation is 10D4d, which is followed immediately by the dimensional oscillation to return to 4D10d as the "dimensional bounce" as shown in Figure 4, which describes the dimensional oscillation from the left to the right: the beginning (inflation as 10D4d through 4D10d to 4D4d), the cosmic expansion-contraction, the end (deflation as 4D4d through 4D10d to 10D4d), the beginning (inflation), the cosmic expansion-contraction, and the end (deflation).

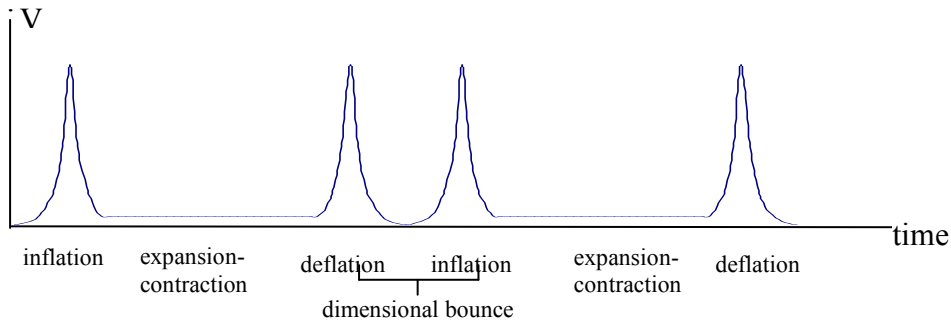

Figure 4. the cyclic observable universe by the dimensional oscillation

The two-step inflation corresponds to the quintom inflation. The symmetry breaking for the light universe can be described by quintom. Quintom [27] [28] [29] is the combination of quintessence and phantom. Quintessence describes a time-varying equation of state parameter, $w$, the ratio of its pressure to energy density and $w > -1$.

$$L_{quintessnec} = \frac{1}{2}(\partial_\mu \phi)^2 - V(\phi) \qquad (25)$$

$$w = \frac{\dot{\phi}^2 - 2V(\phi)}{\dot{\phi}^2 + 2V(\phi)} \qquad (26)$$

$$-1 \leq w \leq +1$$

Quintom includes phantom with $w < -1$. It has opposite sign of kinetic energy.

$$L_{phantom} = \frac{-1}{2}(\partial_\mu \varphi)^2 - V(\varphi) \qquad (27)$$

$$w = \frac{-\dot{\varphi}^2 - 2V(\varphi)}{-\dot{\varphi}^2 + 2V(\varphi)} \qquad (28)$$

$$-1 \geq w$$

As the combination of quintessence and phantom from Eqs. (24), (25), (26), and (27), quintom is as follows.

$$L_{quintessnec} = \frac{1}{2}(\partial_\mu \phi)^2 - \frac{1}{2}(\partial_\mu \varphi)^2 - V(\phi) - V(\varphi) \qquad (29)$$

$$w = \frac{\dot{\phi}^2 - \dot{\varphi}^2 - 2V(\phi) - 2V(\varphi)}{\dot{\phi}^2 - \dot{\varphi}^2 + 2V(\phi) + 2V(\varphi)} \qquad (30)$$

Phantom represents the scalar field $\varphi$ in the space-time dimensional oscillation in QVSL, while quintessence represents the scalar field $\phi$ in the mass dimensional oscillation in the slicing-fusion. Since QVSL does not involve kinetic energy, the physical source of the negative kinetic energy for phantom is the increase in vacuum energy or the emission of the Higgs boson, resulting in the decrease in energy density and pressure with respect to the observable potential, $V(\varphi)$. Combining Eqs. (24c) and (30), quintom is as follows.

$$\begin{aligned} w &= \frac{\dot{\phi}^2 - \dot{\varphi}^2 - 2V(\phi) - 2V(\varphi)}{\dot{\phi}^2 - \dot{\varphi}^2 + 2V(\phi) + 2V(\varphi)} \\ &= \frac{\dot{\phi}^2 - \dot{\varphi}^2 - 2V_{4D10d}(\alpha^{-2\varphi} + \alpha^{2\phi})}{\dot{\phi}^2 - \dot{\varphi}^2 + 2V_{4D10d}(\alpha^{-2\varphi} + \alpha^{2\phi})} \end{aligned} \qquad (31)$$

*where $\varphi \leq 0$ and $\phi \geq 0$*

Figure 5 shows the plot of the evolution of the equation of state *w* for the quintom inflation.

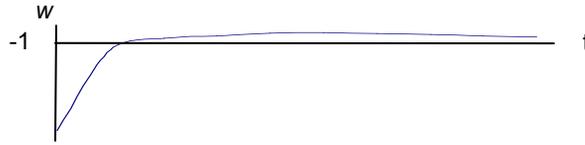

Figure 5. the *w* of quintom for the quintom inflation

Figure 6 shows the plot of the evolution of the equation of state *w* for the cyclic universe as Figure 4.

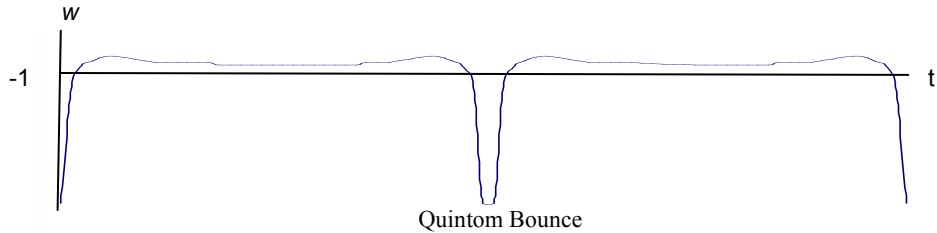

Figure 6. the cyclic universe by the dimensional oscillation as Figure 4

In the dimensional bounce in the middle of Figure 6, the equation of state crosses $w = -1$ twice as also shown in the recent development of the quintom model [30] [31] in which, for the Quintom Bounce, the equation of state crosses the cosmological constant boundary twice around the bounce point to start another cycle of the dual universe.

The hidden dark universe with $D > 4$ and the observable universe with $D = 4$ are the "parallel universes" separated from each other by the bulk space. When the slow QVSL transformation transforms gradually 5D hidden particles in the hidden universe into observable 4D particles, the observable 4 D particles become the dark energy for the observable universe starting from about 5 billion years ago. At a certain time, the hidden universe disappears, and becomes completely observable as dark energy. The maximum connection of the two universes includes the positive energy particle-antiparticle space region, the gravity bulk space region, the negative energy particle-antiparticle space region, and the anti-gravity bulk space region. Through the symmetry among the space regions, all regions expand synchronically and equally. (The symmetry is necessary for the ultimate reversibility of all cosmic processes.) The minimum observable universe has only one of the four space regions before the emergence of dark energy, when the light universe and the dark universe are separated from each other by the bulk space. The present observable universe about reaches the maximum (75%) at the observed 73% dark energy [26], about equal to the three additional space regions to the one original space region. The calculated result from this model at the maximum dark energy gives the universe made up of 75% dark energy, 21.4% dark matter, and 3.6% ordinary matter.

After the maximally connected universe, 4D dark energy transforms back to > 4D particles that are not observable. The removal of dark energy in the observable universe results in the stop of accelerated expansion and the start of contraction of the observable universe.

The end of dark energy starts another "parallel universe period". Both hidden universe and observable universe contract synchronically and equally. Eventually, gravity causes the observable universe to crush to lose all cosmic radiation, resulting in the return to 4D10d particles under the deflation. The increase in vacuum energy allows 4D10d particles to become positive energy 10D4d particles-antiparticle. Meanwhile, hidden > 4D particles-antiparticles in the hidden universe transform into negative energy 10D4d particles-antiparticles. The dual universe can undergo another cycle of the dual universe with the dark and light universes. On the other hand, both universes can undergo transformation by the reverse isodual hole theory to become dual 10D string universe, which in turn can return to the 11D membrane universe as the multiverse background as follows.

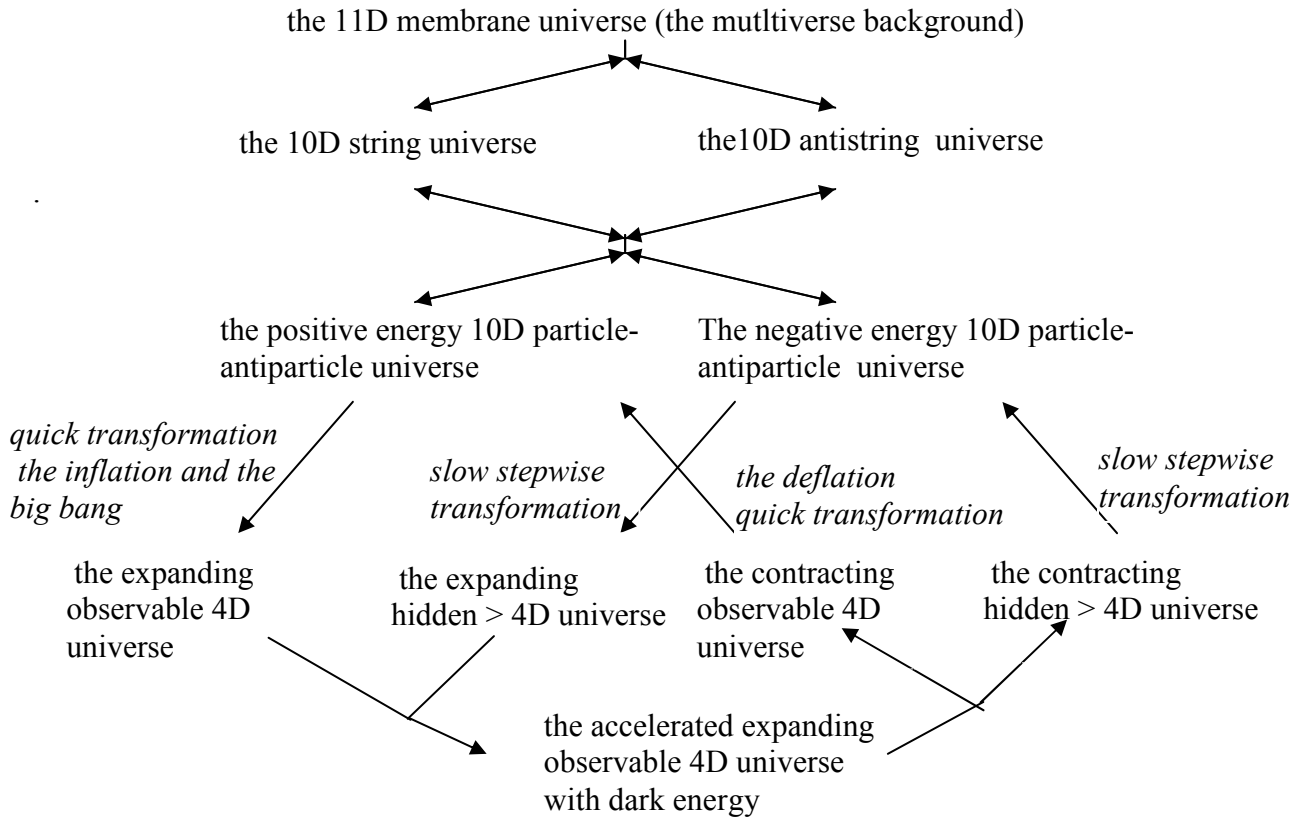

**Figure 7.** Cosmology

## 2.5. Summary

There are three stages of pre-universes in chronological order: the strong pre-universe, the gravitational pre-universe, and the charged pre-universe. The multiverse background is the strong pre-universe with the simplest expression of the cosmic code. Its object structure is 11D membrane and its space structure is attachment space only. The only force is the pre-strong force without gravity. The transformation from 11D membrane to 10D string results in the gravitational pre-universe with both pre-strong force and pre-gravity. The repulsive pre-gravity and pre-antigravity brings about the dual 10D string universe. The coalescence and the separation of the dual universe result in the dual charged universe as dual 10D particle universe with the pre-strong, pre-gravity, and pre-electromagnetic force fields.

The asymmetrical dimensional oscillations result in the asymmetrical dual universe: the light universe with light and kinetic energy and the dark universe without light and kinetic energy. The asymmetrical dimensional oscillation is manifested as the asymmetrical weak force field. The light universe is our observable universe. The dark universe is sometimes hidden, and is sometimes observable as dark energy. The dimensional oscillation for the dark universe is the slow dimensional oscillation from 10D and 4D. The dimensional oscillation for the light universe involves the immediate transformation from 10D to 4D and the introduction of detachment space, resulting in light and kinetic energy.

# 3. The Periodic Table of Elementary Particles

## 3.1. The CP Asymmetry

In the light universe, cosmic radiation is the result of the annihilation of the CP symmetrical particle-antiparticle. However, there is the CP asymmetry, resulting in excess of matter. Matter results from the combination of the CP asymmetrical particle-antiparticle. A baryonic matter particle (4d) has seven dimensional orbitals. The CP asymmetrical particle-antiparticle particle means the combination of two asymmetrical sets of seven from particle and antiparticle, resulting in the combination of the seven "principal dimensional orbitals" and the seven "auxiliary dimensional orbitals". The auxiliary orbitals are dependent on the principal orbitals, so a baryonic matter particle appears to have only one set of dimensional orbitals. For baryonic matter, the principal dimensional orbitals are for leptons and gauge bosons, and the auxiliary dimensional orbitals are mainly for individual quarks. Because of the dependence of the auxiliary dimensional orbitals, individual quarks are hidden. In other words, there is asymmetry between lepton and quark, resulting in the survival of matter without annihilation. The configuration of dimensional orbitals and the periodical table of elementary particles [32] are shown in Fig. 8 and Table 1.

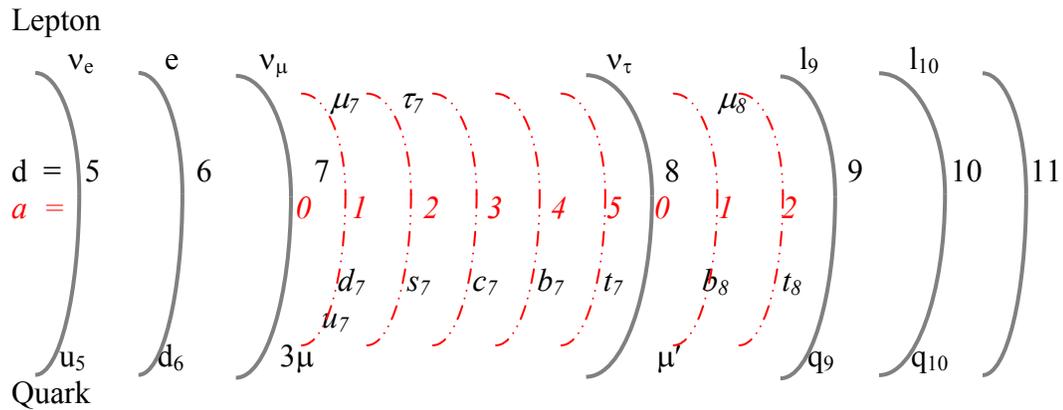

**Fig. 8:** leptons and quarks in the principal and auxiliary dimensional orbitals d = principal dimensional orbital (solid line) number, a = auxiliary dimensional orbital (dot line) number

**Table 1.** The Periodic Table of Elementary Particles
d = principal dimensional orbital number, a = auxiliary dimensional orbital number

| D | a = 0 | 1 | 2 | a = 0 | 1 | 2 | 3 | 4 | 5 | Boson |
|---|---|---|---|---|---|---|---|---|---|---|
|   | Lepton |   |   | Quark |   |   |   |   |   | Boson |
| 5 | $L_5 = \nu_e$ |   |   | $q_5 = u = 3\nu_e$ |   |   |   |   |   | $B_5 = A$ |
| 6 | $L_6 = e$ |   |   | $q_6 = d = 3e$ |   |   |   |   |   | $B_6 = \pi_{1/2}$ |
| 7 | $L_7 = \nu_\mu$ | $\mu_7$ | $\tau_7$ | $q_7 = 3\mu$ | $u_7/d_7$ | $s_7$ | $c_7$ | $b_7$ | $t_7$ | $B_7 = Z_L^0$ |
| 8 | $L_8 = \nu_\tau$ | $\mu_8$ (empty) |   | $q_8 = \mu'$ | $b_8$ | $t_8$ (empty) |   |   |   | $B_8 = X_R$ |
| 9 | $L_9$ |   |   | $q_9$ |   |   |   |   |   | $B_9 = X_L$ |
| 10 |   |   |   |   |   |   |   |   |   | $B_{10} = Z_R^0$ |
| 11 |   |   |   |   |   |   |   |   |   | $B_{11}$ |

In Fig. 8 and Table 1, d is the principal dimensional orbital number, and a is the auxiliary dimensional orbital number. (Note that $F_d$ has lower energy than $B_d$.)

### 3.2. The Boson Mass Formula

The principal dimensional orbitals are for gauge bosons of the force fields. For the gauge bosons, the seven orbitals of principal dimensional orbital are arranged as $F_5$ $B_5$ $F_6$ $B_6$ $F_7$ $B_7$ $F_8$ $B_8$ $F_9$ $B_9$ $F_{10}$ $B_{10}$ $F_{11}$ $B_{11}$, where B and F are boson and fermion in each orbital. The mass dimension in Eq. (17) becomes the orbitals in dimensional orbital with the same equations.

$$M_{d,F} = M_{d,B}\, \alpha_{d,B}, \qquad (32a)$$

$$M_{d-1,B} = M_{d,F}\, \alpha_{d,F}, \qquad (32b)$$

$$M_{d-1,B} = M_{d,B}\, \alpha_d^2. \qquad (32c)$$

where D is the dimensional orbital number from 6 to 11. $E_{5,B}$ and $E_{11,B}$ are the energies for the 5d dimensional orbital and the 11d dimensional orbital, respectively. The lowest energy is the Coulombic field,

$$E_{5,B} = \alpha\, E_{6,F} = \alpha\, M_e. \qquad (33)$$

The bosons generated are the dimensional orbital bosons or $B_D$. Using only $\alpha_e$, the mass of electron ($M_e$), the mass of $Z^0$, and the number (seven) of dimensional orbitals, the masses of $B_D$ as the gauge boson can be calculated as shown in Table 2.

**Table 2.** The Masses of the dimensional orbital bosons:
$\alpha = \alpha_e$, d = dimensional orbital number

| $B_d$ | $M_d$ | GeV (calculated) | Gauge boson | Interaction, symmetry | Predecessor |
|---|---|---|---|---|---|
| $B_5$ | $M_e \alpha$ | $3.7 \times 10^{-6}$ | A | Electromagnetic, U(1) | Pre-charged |
| $B_6$ | $M_e/\alpha$ | $7 \times 10^{-2}$ | $\pi_{1/2}$ | Strong, SU(3) | Pre-strong |
| $B_7$ | $M_6/\alpha_w^2 \cos\theta_w$ | 91.177 (given) | $Z_L^0$ | weak (left), $SU(2)_L$ | Fractionalization (slicing) |
| $B_8$ | $M_7/\alpha^2$ | $1.7 \times 10^6$ | $X_R$ | CP (right) nonconservation | CP asymmetry |
| $B_9$ | $M_8/\alpha^2$ | $3.2 \times 10^{10}$ | $X_L$ | CP (left) nonconservation | CP asymmetry |
| $B_{10}$ | $M_9/\alpha^2$ | $6.0 \times 10^{14}$ | $Z_R^0$ | weak (right) | Fractionalization (slicing) |
| $B_{11}$ | $M_{10}/\alpha^2$ | $1.1 \times 10^{19}$ | G | Gravity | Pregravity |

In Table 2, $\alpha = \alpha_e$ (the fine structure constant for electromagnetic field), and $\alpha_w = \alpha/\sin^2\theta_w$. $\alpha_w$ is not same as $\alpha$ of the rest, because as shown later, there is a mixing between $B_5$ and $B_7$ as the symmetry mixing between U(1) and SU(2) in the standard theory of the electroweak interaction, and $\sin\theta_w$ is not equal to 1. (The symmetrical charged dual pre-universe overlaps with the current asymmetrical universe for the weak interaction as shown earlier.) As shown later, $B_5$, $B_6$, $B_7$, $B_8$, $B_9$, and $B_{10}$ are A (massless photon), $\pi_{1/2}$ (half of pion), $Z_L^0$, $X_R$, $X_L$, and $Z_R^0$, respectively, responsible for the electromagnetic field, the strong interaction, the weak (left handed) interaction, the CP (right handed) nonconservation, the CP (left handed) nonconservation, and the P (right handed) nonconservation, respectively. The calculated value for $\alpha_w$ is 0.2973, and $\theta_w$ is $29.69^0$ in good agreement with $29.31^0$ for the observed value of $\theta_w$ [33]. The calculated energy for $B_{11}$ is $1.1 \times 10^{19}$ GeV in good agreement with the Planck mass, $1.2 \times 10^{19}$ GeV. The strong interaction, representing by $\pi_{1/2}$ (half of pion), is for the interactions among quarks, and for the hiding of individual quarks in the auxiliary orbital. The weak interaction, representing by $Z_L^0$, is for the interaction involving changing flavors (decomposition and condensation) among quarks and leptons.

There are dualities between dimensional orbitals and the cosmic evolution process. The pre-charged force, the pre-strong force, the fractionalization, the CP asymmetry, and the pregravity are the predecessors of electromagnetic force, the strong force, the weak interaction, the CP nonconservation, and gravity, respectively. These forces are manifested in the dimensional orbitals with various space-time symmetries and gauge symmetries. The strengths of these forces are different than their predecessors, and are arranged according to the dimensional orbitals. Only the 4d particle (baryonic matter) has the $B_5$, so without $B_5$, dark matter consists of permanently neutral higher dimensional particles. It cannot emit light, cannot form atoms, and exists as neutral gas.

The principal dimensional boson, $B_8$, is a CP violating boson, because $B_8$ is assumed to have the CP-violating $U(1)_R$ symmetry. The ratio of the force constants between the CP-invariant $W_L$ in $B_8$ and the CP-violating $X_R$ in $B_8$ is

$$\begin{aligned}\frac{G_8}{G_7} &= \frac{\alpha\, E_7^2\, \cos^2\Theta_W}{\alpha_W\, E_8^2} \\ &= 5.3 \times 10^{-10},\end{aligned} \qquad (34)$$

which is in the same order as the ratio of the force constants between the CP-invariant weak interaction and the CP-violating interaction with $|\Delta S| = 2$.

The principal dimensional boson, $B_9$ ($X_L$), has the CP-violating U(1)$_L$ symmetry. $B_9$ generates matter. The ratio of force constants between $X_R$ with CP conservation and $X_L$ with CP-nonconservation is

$$\begin{aligned}\frac{G_9}{G_8} &= \frac{\alpha\, E_8^2}{\alpha\, E_9^2} \\ &= \alpha^4 \\ &= 2.8 \times 10^{-9},\end{aligned} \qquad (35)$$

which is the ratio of the numbers between matter (dark and baryonic) and photons in the universe. It is close to the ratio of the numbers between baryonic matter and photons about $5 \times 10^{-10}$ obtained by the big bang nucleosynthesis.

Auxiliary dimensional orbital is derived from principal dimensional orbital. It is for high-mass leptons and individual quarks. Auxiliary dimensional orbital is the second set of the three sets of seven orbitals. The combination of dimensional auxiliary dimensional orbitals constitutes the periodic table for elementary particles as shown in Fig. 8 and Table 1.

There are two types of fermions in the periodic table of elementary particles: low-mass leptons and high-mass leptons and quarks. Low-mass leptons include $\nu_e$, $e$, $\nu_\mu$, and $\nu_\tau$, which are in principal dimensional orbital, not in auxiliary dimensional orbital. $l_d$ is denoted as lepton with principal dimension number, $d$. $l_5$, $l_6$, $l_7$, and $l_8$ are $\nu_e$, $e$, $\nu_\mu$, and $\nu_\tau$, respectively. All neutrinos have zero mass because of chiral symmetry (permanent chiral symmetry).

### 3.3. The Mass Composites of Leptons and Quarks

High-mass leptons and quarks include $\mu$, $\tau$, u, d, s, c, b, and t, which are the combinations of both principal dimensional fermions and auxiliary dimensional fermions. Each fermion can be defined by principal dimensional orbital numbers (d's) and auxiliary dimensional orbital numbers (a's) as $d_a$ in Table 3. For examples, e is $6_0$ that means it has d (principal dimensional orbital number) = 6 and a (auxiliary dimensional orbital number) = 0, so e is a principal dimensional fermion.

High-mass leptons, $\mu$ and $\tau$, are the combinations of principal dimensional fermions, e and $\nu_\mu$, and auxiliary dimensional fermions. For example, $\mu$ is the combination of e, $\nu_\mu$, and $\mu_7$, which is $7_1$ that has d = 7 and a = 1.

Quarks are the combination of principal dimensional quarks ($q_d$) and auxiliary dimensional quarks. The principal dimensional fermion for quark is derived from principal dimensional lepton. To generate a principal dimensional quark in principal dimensional orbital from a lepton in the same principal dimensional orbital is to add the lepton to the boson from the combined lepton-antilepton. Thus, the mass of the quark is three times of the mass of the corresponding lepton in the same dimension. The equation for the mass of principal dimensional fermion for quark is

$$M_{q_d} = 3M_{l_d} \tag{36}$$

For principal dimensional quarks, $q_5$ ($5_0$) and $q_6$ ($6_0$) are $3\nu_e$ and $3e$, respectively. Since $l_7$ is massless $\nu_\mu$, $\nu_\mu$ is replaced by $\mu$, and $q_7$ is $3\mu$. Quarks are the combinations of principal dimensional quarks, $q_d$, and auxiliary dimensional quarks. For example, s quark is the combination of $q_6$ (3e), $q_7$ ($3\mu$) and $s_7$ (auxiliary dimensional quark = $7_2$).

Each fermion can be defined by principal dimensional orbital numbers (d's) and auxiliary dimensional orbital numbers (a's). All leptons and quarks with d's, a's and the calculated masses are listed in Table 3.

**Table 3.** The Compositions and the Constituent Masses of Leptons and Quarks
d = principal dimensional orbital number and a = auxiliary dimensional orbital number

| | $d_a$ | Composition | Calculated Mass |
|---|---|---|---|
| Leptons | $d_a$ for leptons | | |
| $\nu_e$ | $5_0$ | $\nu_e$ | 0 |
| E | $6_0$ | e | 0.51 MeV (given) |
| $\nu_\mu$ | $7_0$ | $\nu_\mu$ | 0 |
| $\nu_\tau$ | $8_0$ | $\nu_\tau$ | 0 |
| $\mu$ | $6_0 + 7_0 + 7_1$ | $e + \nu_\mu + \mu_7$ | 105.6 MeV |
| $\tau$ | $6_0 + 7_0 + 7_2$ | $e + \nu_\mu + \tau_7$ | 1786 MeV |
| $\mu'$ | $6_0 + 7_0 + 7_2 + 8_0 + 8_1$ | $e + \nu_\mu + \mu_7 + \nu_\tau + \mu_8$ | 136.9 GeV |
| Quarks | $d_a$ for quarks | | |
| U | $5_0 + 7_0 + 7_1$ | $q_5 + q_7 + u_7$ | 330.8 MeV |
| D | $6_0 + 7_0 + 7_1$ | $q_6 + q_7 + d_7$ | 332.3 MeV |
| S | $6_0 + 7_0 + 7_2$ | $q_6 + q_7 + s_7$ | 558 MeV |
| C | $5_0 + 7_0 + 7_3$ | $q_5 + q_7 + c_7$ | 1701 MeV |
| B | $6_0 + 7_0 + 7_4$ | $q_6 + q_7 + b_7$ | 5318 MeV |
| T | $5_0 + 7_0 + 7_5 + 8_0 + 8_2$ | $q_5 + q_7 + t_7 + q_8 + t_8$ | 176.5 GeV |

### 3.4. The Lepton Formula

The principal dimensional fermion for heavy leptons ($\mu$ and $\tau$) is e and $\nu_e$. Auxiliary dimensional fermion is derived from principal dimensional boson in the same way as Eq. (32) to relate the energies for fermion and boson. For the mass of auxiliary dimensional fermion (AF) from principal dimensional boson (B), the equation is Eq. (37).

$$M_{AF_{d,a}} = \frac{M_{B_{d-1,0}}}{\alpha_a} \sum_{a=0}^{a} a^4 \quad, \tag{37}$$

where $\alpha_a$ = auxiliary dimensional fine structure constant, and a = auxiliary dimension number = 0 or integer. The first term, $\dfrac{M_{B_{D-1,0}}}{\alpha_a}$, of the mass formula (Eq.(37)) for the auxiliary dimensional fermions is derived from the mass equation, Eq. (32), for the principal dimensional fermions and bosons. The second term, $\sum_{a=0}^{a} a^4$, of the mass formula is for Bohr-Sommerfeld quantization for a charge - dipole interaction in a circular orbit as described by A. Barut [34]. As in Barut lepton mass formula, $1/\alpha_a$ is 3/2. The coefficient, 3/2, is to convert the principal dimensional boson mass to the mass of the auxiliary dimensional fermion in the higher dimension by adding the boson mass to its fermion mass which is one-half of the boson mass. Using Eq. (32), Eq. (37) becomes the formula for the mass of auxiliary dimensional fermions (AF).

$$M_{AF_{d,a}} = \dfrac{3 M_{B_{d-1,0}}}{2} \sum_{a=0}^{a} a^4$$

$$= \dfrac{3 M_{F_{d-1,0}}}{2\alpha_{d-1}} \sum_{a=0}^{a} a^4 \qquad (38)$$

$$= \dfrac{3}{2} M_{F_{d,0}} \alpha_d \sum_{a=0}^{a} a^4$$

The mass of this auxiliary dimensional fermion is added to the sum of masses from the corresponding principal dimensional fermions (F's) with the same electric charge or the same dimension. The corresponding principal dimensional leptons for u (2/3 charge) and d (-1/3 charge) are $\nu_e$ (0 charge) and e (-1 charge), respectively, by adding –2/3 charge to the charges of u and d [35]. The fermion mass formula for heavy leptons is derived as follows.

$$M_{F_{d,a}} = \sum M_F + M_{AF_{d,a}}$$

$$= \sum M_F + \dfrac{3 M_{B_{d-1,0}}}{2} \sum_{a=0}^{a} a^4 \qquad (39a)$$

$$= \sum M_F + \dfrac{3 M_{F_{d-1,0}}}{2\alpha_{d-1}} \sum_{a=0}^{a} a^4 \qquad (39b)$$

$$= \sum M_F + \dfrac{3}{2} M_{F_{d,0}} \alpha_d \sum_{a=0}^{a} a^4 \qquad (39c)$$

Eq. (39b) is for the calculations of the masses of leptons. The principal dimensional fermion in the first term is e. Eq. (39b) can be rewritten as Eq. (40).

$$M_a = M_e + \frac{3M_e}{2\alpha} \sum_{a=0}^{a} a^4, \qquad (40)$$

a = 0, 1, and 2 are for e, μ, and τ, respectively. It is identical to the Barut lepton mass formula.

### 3.5. The Quark Mass Formula

The auxiliary dimensional quarks except a part of t quark are $q_7$'s. Eq.(39c) is used to calculate the masses of quarks. The principal dimensional quarks include $3\nu_\mu$, 3e, and 3μ., $\alpha_7 = \alpha_w$, and $q_7 = 3\mu$. Eq. (39c) can be rewritten as the quark mass formula.

$$M_q = \sum M_F + \frac{3\alpha_w M_{3\mu}}{2} \sum_{a=0}^{a} a^4, \qquad (41)$$

where a = 1, 2, 3, 4, and 5 for u/d, s, c, b, and a part of t, respectively.

To match $l_8$ ($\nu_\tau$), quarks include $q_8$ as a part of t quark. In the same way that $q_7 = 3\mu$, $q_8$ involves μ'. μ' is the sum of e, μ, and $\mu_8$ (auxiliary dimensional lepton). Using Eq. (39a), the mass of $\mu_8$ is equal to 3/2 of the mass of $B_7$, which is $Z^0$. Because there are only three families for leptons, μ' is the extra lepton, which is "hidden". μ' can appear only as μ + photon. The pairing of μ + μ from the hidden μ' and regular μ may account for the occurrence of same sign dilepton in the high energy level [36]. The principal dimensional quark $q_8 = \mu'$ instead of 3μ', because μ' is hidden, and $q_8$ does not need to be 3μ' to be different. Using the equation similar to Eq.(41), the calculation for t quark involves $\alpha_8 = \alpha$, μ' instead of 3μ for principal fermion, and a = 1 and 2 for $b_8$ and $t_8$, respectively. The hiding of μ' for leptons is balanced by the hiding of $b_8$ for quarks.

The calculated masses are in good agreement with the observed constituent masses of leptons and quarks [37]. The mass of the top quark [38] is 174.3 ± 5.1 GeV in a good agreement with the calculated value, 176.5 GeV.

With the masses of quarks calculated by the periodic table of elementary particles, the masses of all hardrons can be calculated [32] as the composes of quarks, as molecules are the composes of atoms. The calculated values are in good agreement with the observed values. For examples, the calculated masses of neutron and pion are 939.54MeV, and 135.01MeV in excellent agreement with the observed masses, 939.57 MeV and 134.98 MeV, respectively. At different temperatures, the strong force (QCD) among quarks in hadrons behaves differently to follow different dimensional orbitals [33].

### 3.6. Summary

For baryonic matter, the incorporation of detachment space for baryonic matter brings about "the dimensional orbitals" as the base for the periodic table of elementary particles for all leptons, quarks, and gauge bosons. The masses of gauge bosons, leptons, quarks can be calculated using only four known constants: the number of the extra spatial dimensions in the eleven-dimensional membrane, the mass of electron, the mass of Z°, and the fine structure constant. The calculated values are in good agreement with the observed values. The differences in dimensional orbitals result in incompatible dark matter and baryonic matter.

# 4. The Galaxy Formation

## Introduction

The current observable universe contains dark energy, dark matter, and baryonic matter. As mentioned in the previous section, dark energy is from the dark universe to accelerate the expansion of the observable universe. Dark matter have different mass dimension from the baryonic matter. We live in the world of baryonic matter. The separation of baryonic matter and dark matter results in the galaxy formation.

## 4.1. The Separation of Baryonic Matter and Dark Matter

Dark matter has been detected only indirectly by means of its gravitational effects astronomically. Dark matter as weakly interacting massive particles (WIMPs) has not been detected directly on the earth [39]. The previous section proposes that the absence of the direct detection of dark matter on the earth is due to the incompatibility between baryonic matter and dark matter, analogous to incompatible water and oil. The previous section provides the reasons for the incompatibility and the mass ratio (6 to 1) of dark matter to baryonic matter. Basically, during the inflation before the big bang, dark matter, baryonic matter, cosmic radiation, and the gauge force fields are generated. There are six types of dark matter with the "mass dimensions' from 5 to 10, while baryonic matter has the mass dimension of 4. As a result, the mass ratio is 6 to 1 as observed. Without electromagnetism, dark matter cannot emit light, and is incompatible to baryonic matter. Like oil, dark matter is completely non-polar. The common link between baryonic matter and dark matter is the cosmic radiation resulted from the annihilation of matter and antimatter from both baryonic matter and dark matter. The cosmic radiation is coupled strongly to baryonic matter through the electromagnetism, and weakly to dark matter without electromagnetism. With the high concentration of cosmic radiation at the beginning of the big bang, baryonic matter and dark matter are completely compatible. As the universe ages and expands, the concentration of cosmic concentration decreases, resulting in the increasing incompatibility between baryonic matter and dark matter until the incompatibility reaches to the maximum value with low concentration of cosmic radiation.

The incompatibility is expressed in the form of the repulsive MOND (modified Newtonian dynamics) force field. MOND [40] proposes the deviation from the Newtonian dynamics in the low acceleration region in the outer region of a galaxy. This paper proposes the MOND forces in the interface between the baryonic matter region and the dark matter region [41]. In the interface, the same matter materials attract as the conventional attractive MOND force, and the different matter materials repulse as the repulsive MOND force between baryonic matter and dark matter.

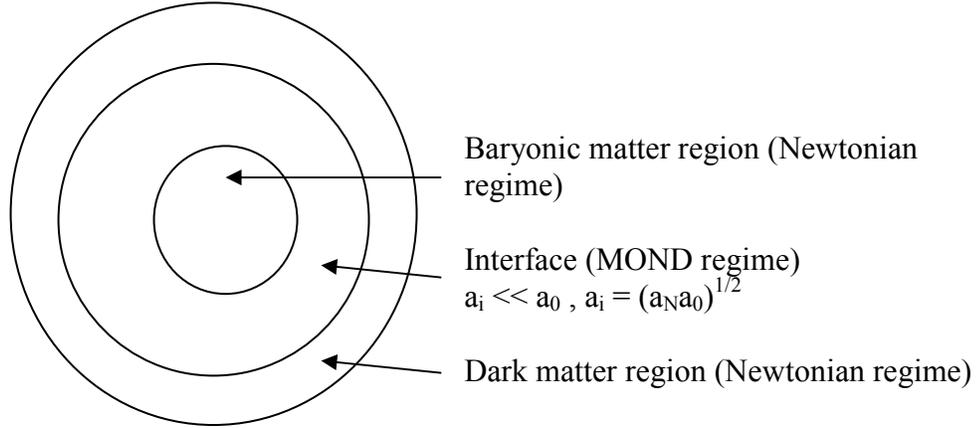

Baryonic matter region (Newtonian regime)

Interface (MOND regime)
$a_i \ll a_0$, $a_i = (a_N a_0)^{1/2}$

Dark matter region (Newtonian regime)

Figure 9: the interfacial region between the baryonic and the dark matter regions

In Figure 9, the inner part is the baryonic matter region, the middle part is the interface, and the outer part is the dark matter region. The MOND forces in the interface are the interfacial attractive force (conventional MOND force), $F_{i-A}$, among the same matter materials and the interfacial repulsive force (repulsive MOND force), $F_{i-R}$, between baryonic matter material and dark matter material. The interfacial repulsive force enhances the interfacial attractive force toward the center of gravity in terms of the interfacial acceleration, $a_i$.

The border between the baryonic matter region and the interface is defined by the acceleration constant, $a_0$. The interfacial acceleration is less than $a_0$. The enhancement is expressed as the square root of the product of $a_i$ and $a_0$. In the baryonic matter region, $a_b$ is greater than $a_0$, and is equal to normal Newtonian acceleration as Eq. (42).

$$\begin{aligned} a_0 \ll a_b, \quad a_b = a_N \quad \text{in the baryonic matter region} \\ a_0 \gg a_i, \quad a_i = \sqrt{a_N a_0} \quad \text{in the interfacial region} \end{aligned} \quad (42)$$

The interfacial attractive force in the interface with the baryonic matter region is expressed as Eq. (43) where m is the mass of baryonic material in the interface.

$$\begin{aligned} F_{i-A} &= m a_N \\ &= m \frac{a_i^2}{a_O}, \end{aligned} \quad (43)$$

The comparison of the interfacial attractive force, $F_{i-A}$, and the non-existing interfacial Newtonian attractive force, $F_{i-Newton}$ in the interface is as Eqs. (44), (45), and (46), where G is the gravitation constant, M is the mass of the baryonic material, and r the distance between the gravitational center and the material in the infacial region.

$$F_{i-A} = \frac{GMm}{r^2}$$

$$= m\frac{a^2}{a_0}, \tag{44}$$

$$F_{i-Newton} = \frac{GMm}{r^2}$$

$$= ma$$

$$a_i = \frac{\sqrt{GMa_0}}{r} \tag{45}$$

$$a_{i-Newron} = \frac{GM}{r^2},$$

$$F_{i-A} = \frac{m\sqrt{GMa_0}}{r} \tag{46}$$

$$F_{i-Newron} = \frac{mGM}{r^2},$$

The interfacial attractive force decays with r, while the interfacial Newtonian force decays with $r^2$. Therefore, in the interface when $a_0 \gg a_i$, with sufficient dark matter, the interfacial repulsive force, $F_{i-R}$, is the difference between the interfacial attractive force and the interfacial Newtonian force as Eq. (47).

$$a_0 \gg a_i, \text{ in the interfacial region}$$
$$F_{i-R} = F_{i-A} - F_{i-Newton} \tag{47}$$
$$= m(\frac{\sqrt{GMa_0}}{r} - \frac{GM}{r^2})$$

The same interfacial attractive force and the interfacial repulsive force also occur for dark matter in the opposite direction. Thus, the repulsive MOND force filed results in the separation of baryonic matter and dark matter.

The acceleration constant, $a_0$, represents the maximum acceleration constant for the maximum incompatibility between baryonic matter and dark matter. The common link between baryonic matter and dark matter is cosmic radiation resulted from the annihilation of matter and antimatter from both baryonic matter and dark matter. With the high concentration of cosmic radiation at the big bang, baryonic matter and dark matter are completely compatible. As the universe ages and expands, the concentration of cosmic concentration decreases, resulting in the increasing incompatibility between baryonic matter and dark matter. The incompatibility reaches maximum when the concentration of cosmic radiation becomes is too low for the compatibility between baryonic matter and dark matter. Therefore, for the early universe before the formation of galaxy when the concentration of cosmic radiation is still high, the time-dependent Eq. (42) is as Eq. (48).

$$a_i = \sqrt{\frac{a_N a_0 t}{t_0}} \quad for\ t_0 \geq t, \tag{48}$$

where t is the age of the universe, and $t_0$ is the age of the universe to reach the maximum incompatibility between baryonic matter and dark matter.

The distance, $r_0$, from the center to the border of the interface is as Eq. (49).

$$r_0 = \sqrt{GM/a_0} \tag{49}$$

In the early universe, $r_0$ decreases with the age of the universe as Eq. (50).

$$r_0 = \sqrt{\frac{GM t_0}{a_0 t}} \tag{50}$$

The decreases in $r_0$ leads to the increase in the interface where the interfacial forces exist. The interfacial forces also increase with time.

$$a_0 \gg a_i,\ in\ the\ \mathrm{int}erfacial\ region$$
$$F_{i-R} = F_{i-A} - F_{i-Newton} \tag{51}$$
$$= m(\frac{\sqrt{GM a_0 t/t_0}}{r} - \frac{GM}{r^2})$$

To minimize the interface and the interfacial forces, the same matter materials increasingly come together to form the matter droplets separating from the different matter materials. The increasing formation of the matter droplets with increasing incompatibility is similar to the increasing formation of oil droplets with increasing incompatibility between oil and water. Since there are more dark matter materials than baryonic matter materials, most of the matter droplets are baryonic droplets surrounded by dark matter materials. The early universe is characterized by the increases in the size and the number of the matter droplets due to the increasing incompatibility between baryonic matter and dark matter.

### 4.2. The Formation of the Inhomogeneous Structures

The Inflationary Universe scenario [42] provides possible solutions of the horizon, flatness and formation of structure problems. In the standard inflation theory, quantum fluctuations during the inflation are stretched exponentially so that they can become the seeds for the formation of inhomogeneous structure such as galaxies and galaxy clusters.

This paper posits that the inhomogeneous structure comes from both quantum fluctuation during the inflation and the repulsive MOND force between baryonic matter and dark matter after the inflation. As mentioned in the previous section, the increasing repulsive MOND force field with the increasing incompatibility in the early universe results in the increase in the size and number of the matter droplets.

For the first few hundred thousand years after the Big Bang (which took place about 13.7 billion years ago), the universe was a hot, murky mess, with no light radiating out. Because there is no residual light from that early epoch, scientists can't observe any traces of it. But about 400,000 years after the Big Bang, temperatures in the universe cooled, electrons and protons joined to form neutral hydrogen as the recombination. The inhomogeneous structure as the baryonic droplets by the incompatibility between baryonic matter and dark matter is observed [43] as anisotropies in CMB (cosmic microwave background).

As the universe expanded after the time of recombination, the density of cosmic radiation decreases, and the size of the baryonic droplets increased with the increasing incompatibility between baryonic matter and dark matter. The growth of the baryonic droplet by the increasing incompatibility from the cosmic expansion coincided with the growth of the baryonic droplet by gravitational instability from the cosmic expansion. The formation of galaxies is through both gravitational instability and the incompatibility between baryonic matter and dark matter.

The pre-galactic universe consisted of the growing baryonic droplets surrounded by the dark matter halos, which connected among one another in the form of filaments and voids. These dark matter domains later became the dark matter halos, and the baryonic droplets became galaxies, clusters, and superclusters.

When there were many baryonic droplets, the merger among the baryonic droplets became another mechanism to increase the droplet size and mass. When three or more homogeneous baryonic droplets merged together, dark matter was likely trapped in the merged droplet (C, D, E, and F in Fig. 10). The droplet with trapped dark matter inside is the heterogeneous baryonic droplet, while the droplet without trapped dark matter inside is the homogeneous baryonic droplet.

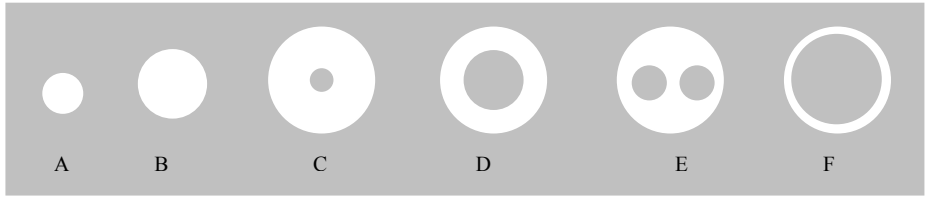

**Fig. 10**: the homogeneous baryonic droplets (A, and B), and the heterogeneous baryonic droplets (C, D, E, and F)

In the heterogeneous droplets C, D, E, and F, dark matter was trapped in the cores of the baryonic droplets. Because of the prevalence of dark matter, almost all baryonic droplets were the heterogeneous droplets. There were the dark matter core, the baryonic matter shell, and the dark matter halo around the baryonic droplet, resulting in two repulsive forces as the pressures between the dark matter core and the baryonic matter shell and between the baryonic shell and the dark matter halo. In the equilibrium state, the internal pressure between the dark matter core and the baryonic matter shell was same as the external pressure between the baryonic shell and the dark matter halo.

When the temperature dropped to ~ 1000°K, some hydrogen atoms in the droplet paired up to create the primordial molecular layers. Molecular hydrogen cooled the primordial molecular layers by emitting infrared radiation after collision with atomic hydrogen. Eventually, the temperature of the molecular layers dropped to around 200 to 300°K, reducing the gas pressure and allowing the molecular layers to continue contracting into gravitationally bound dense primordial molecular clouds. The diameters of the primordial could be up to 100 light-years with the masses of up to 6 million solar masses. Most of baryonic droplets contained thousands of the primordial molecular clouds.

The formation of the primordial molecular clouds created the gap in the baryonic matter shell. The gap allowed the dark matter in the dark matter core to leak out, resulting in a tunnel between the dark matter core and the external dark matter halo. The continuous leaking of the dark matter expanded the tunnel. Consequently, the dark matter in the dark matter core rushed out of the dark matter core, resulting in the "big eruption". The ejection of the dark matter from the dark matter core reduced the internal pressure between the dark matter core and the baryonic matter shell. The external pressure between the baryonic matter shell and the dark matter halo caused the collapse of the baryonic droplet. The collapse of the baryonic droplet is like the collapse of a balloon as the air (as dark matter) moves out the balloon.

The collapse of the baryonic droplet forced the head-on collisions of the primordial molecular clouds in the baryonic matter shell. In the center of the collapsed baryonic droplet, the head-on collisions of the primordial molecular clouds generated the shock wave as the turbulence in the collided primordial molecular clouds. The turbulence triggered the collapse of the core of the primordial cloud. The core fragmented into multiple stellar embryos, in each a protostar nucleated and pulled in gas. Without the heavy elements to dissipate heat, the mass of the primordial protostar was 500 to 1,000 solar masses at about 200°K. The primordial protostar shrank in size, increased in density, and became the primordial massive star when nuclear fusion began in its core. The massive primordial star formation is as follows.

$$\text{incompatible dark matter and baryonic matter} \longrightarrow \text{homogeneous baryonic droplets} \xrightarrow{combination} \text{heterogeneous baryonic droplet} \xrightarrow{the\ cooling} \text{molecular clouds in baryonic matter shell} \xrightarrow{eruption,\ collapse,\ and\ collision} \text{protostar} \xrightarrow{nuclear\ fusion} \text{massive primordial star}$$

The intense UV radiation from the high surface temperature of the massive primordial stars started the reionization effectively, and also triggered further star formation. The massive primordial stars were short-lived (few million years old). The explosion of the massive primordial stars was the massive supernova that caused reionization and triggered star formation. The heavy elements generated during the primordial star formation scattered throughout the space. The dissipation of heat by heavy elements allowed the normal rather than massive star formation. With many ways to trigger star formation, the rate of star formation increased rapidly. The big eruption that initiated the star formation started to occur about 400 million years after the big bang, and the reionization started to occur soon after. The rate of star formation peaked about 2 billion years after the big bang [44].

Since the head-on collision of the molecular clouds took place at the center of the collapsed baryonic droplet, the star formation started in the center of the collapsed baryonic droplet. With other ways to trigger star formation, the star formation propagated away from the center. The star formation started from the center from which the star formation propagated, so the primordial galaxies appeared to be small surrounded by the large hydrogen blobs. The surrounding large hydrogen blobs corresponds to the observed Lyman alpha blobs of Lyman alpha (Lyα) emission by hydrogen, which have been discovered in the vicinity of galaxies at early cosmic times. The amount of hydrogen in the blobs was also increased by the incoming abundant intergalactic hydrogen. The repulsive dark matter halos prevented the hydrogen gas inside from escaping from the galaxies. Dijkstra and Loeb [45] posited that the early galaxies grew quickly by the cold accretion mode from the observed Lyman alpha blobs. The growth by the merger of galaxies was too slow for the observed fast growth of the early galaxies.

If there was small dark matter core as in the heterogeneous baryonic droplet (C in Figure 10), the big eruption took relatively short time to cause the collapse of the baryonic droplet. The change in the shape of the baryonic droplet after the collapse was relatively minor. The collapse results in elliptical shape in $E_0$ to $E_7$ elliptical galaxies, whose lengths of major axes are proportional to the relative sizes of the dark matter core. Because of the short time for the collapse of the baryonic droplet, the star formation by the collapse occurred quickly at the center.

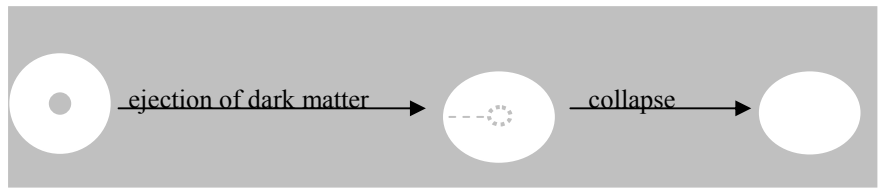

Most of the primordial stars merged to form the supermassive center, resulting in the quasar galaxies. Such first quasar galaxies that occurred as early as z = 6.28 were observed to have about the same sizes as the Milky Way [46]. This formation of galaxy follows the monolithic collapse model in which baryonic gas in galaxies collapses to form stars within a very short period, so there are small numbers of observed young stars in elliptical galaxies. Elliptical galaxies continue to grow slowly as the universe expands.

If the size of the dark matter core is medium (D in Fig. 10), the collapse of the baryonic droplet caused a large change in shape, resulting in the rapidly rotating disk as spiral galaxy. The rapidly rotating disk underwent differential rotation with the increasing angular speeds toward the center. After few rotations, the structure consisted of a bungle was formed and the attached spiral arms as spiral galaxy as Fig. 11.

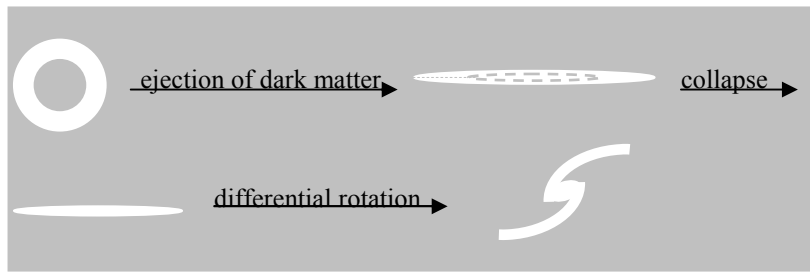

**Fig. 11:** the formation of spiral galaxy

The spiral galaxy took longer time to erupt and collapse than the elliptical galaxy, so the star formation was later than elliptical galaxy. Because of the large size of the dark matter core, the density of the primordial molecular clouds was lower than elliptical galaxy, so the rate of star formation in spiral galaxy is slower than elliptical galaxy. During the collapse of the baryonic droplet, some primordial molecular clouds moved away to form globular clusters near the main group of the primordial molecular clouds. Most of the primordial massive stars merged to form the supermassive center. The merge of spiral galaxies with comparable sizes destroys the disk shape, so most spiral galaxies are not merged galaxies.

When two dark matter cores inside far apart from each other (E in Fig. 10) generated two openings in opposite sides of the droplet, the dark matter could eject from both openings. The two opening is equivalent to the overlapping of two ellipses, resulting in the thick middle part, resulting in the star formation in the thick middle part and the formation of barred spiral galaxy. The differential rotation is similar to that of spiral galaxy as Fig. 12.

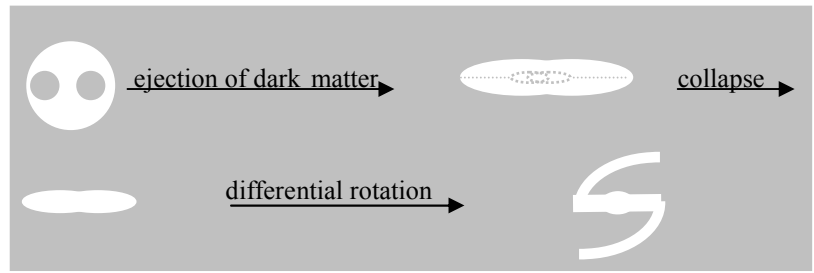

**Fig. 12**: the formation of barred spiral galaxy

As in normal spiral galaxy, the length of the spiral arm depends on the size of the dark matter core. The smallest dark matter core for barred spiral galaxy brings about SBa, and the largest dark matter core brings about SBd. The stars form in the low-density spiral arms much later than in the nucleus, so they are many young stars in the spiral arms. In barred spiral galaxy, because of the larger dark matter core area than normal spiral galaxy, the star formation occurred later than normal spiral galaxy, and the rate of star formation was slower than normal spiral galaxy.

If the size of the dark matter core was large (F in Fig. 10), the eruption of the dark matter in the dark matter core occurred easily in multiple places. The baryonic matter shell became fragmented, resulting in irregular galaxy. The turbulence from the collapse of the baryonic droplet was weak, and the density of the primordial molecular clouds was low, so the rate of star formation was slow. The star formation continues in a slow rate up to the present time.

At the end of the big eruption, vast majority of baryonic matter was primordial free baryonic matter resided in dark matter outside of the galaxies from the big eruption. This free baryonic matter constituted the intergalactic medium (IGM). Stellar winds, supernova winds, and quasars provide heat and heavy elements to the IGM as ionized baryonic atoms. The heat prevented the formation of the baryonic droplet in the IGM.

Galaxies merged into new large galaxies, such as giant elliptical galaxy and cD galaxy (z > 1-2). Similar to the transient molecular cloud formation from the ISM (inter-stellar medium) through turbulence, the tidal debris and turbulence from the mergers generated the numerous transient molecular regions, which located in a broad area [47].

The incompatibility between baryonic matter and dark matter transformed these transient molecular regions into the stable second-generation baryonic droplets surrounded by the dark matter halos. The baryonic droplets had much higher fraction of hydrogen molecules, much lower fraction of dark matter, higher density, and lower temperature, and lower entropy than the surrounding.

During this period, the acceleration constant reached to the maximum value with the maximum incompatibility between baryonic matter and dark matter. The growth of the baryonic droplets did not depend on the increasing incompatibility. The growth of the baryonic droplets depended on the turbulences that carried IGM to the baryonic droplets. The rapid growth of the baryonic droplets drew large amount of the surrounding IGM inward, generating the IGM flow shown as the cooling flow. The IGM flow induced the galaxy flow. The IGM flow and the galaxy flow moved toward the merged galaxies, resulting in the protocluster (z ~ 0.5) with the merged galaxies as the cluster center.

Before the protocluster stage, spirals grew normally and passively by absorbing gas from the IGM as the universe expanded. During the protoculster stage (z ~ 0.5), the massive IGM flow injected a large amount of gas into the spirals that joined in the galaxy flow. Most of the injected hot gas passed through the spiral arms and settled in the bungle parts of the spirals. Such surges of gas absorption from the IGM flow resulted in major starbursts (z ~ 0.4) [48]. Meanwhile, the nearby baryonic droplets continued to draw the IGM, and the IGM flow and the galaxy flow continued. The results were the formation of high-density region, where the galaxies and the baryonic droplets competed for the IGM as the gas reservoir. Eventually, the maturity of the baryonic droplets caused a decrease in drawing the IGM inward, resulting in the slow IGM flow. Subsequently, the depleted gas reservoir could not support the major starbursts (z ~ 0.3). The galaxy harassment and the mergers in this high-density region disrupted the spiral arms of spirals, resulting in S0 galaxies with indistinct spiral arms (z ~ 0.1 – 0.25). The transformation process of spirals into S0 galaxies started at the core first, and moved to the outside of the core. Thus, the fraction of spirals decreases with decreasing distance from the cluster center.

The static and slow-moving second-generation baryonic droplets turned into dwarf elliptical galaxies and globular clusters. The fast moving second-generation baryonic droplets formed the second-generation baryonic stream, which underwent a differential rotation to minimize the interfacial area between the baryonic matter and dark matter. The result is the formation of blue compact dwarf galaxies (BCD), such as NGC 2915 with very extended spiral arms. Since the star formation is steady and slow, so the stars formed in BCD are new.

The galaxies formed during z < 0.1-0.2 are mostly metal-rich tidal dwarf galaxies (TDG) from tidal tails torn out from interacting galaxies. In some cases, the tidal tail and the baryonic droplet merge to generate the starbursts with higher fraction of molecule than the TDG formed by tidal tail alone [49].

When the interactions among large galaxies were mild, the mild turbulence caused the formation of few molecular regions, which located in narrow area close to the large galaxies. Such few molecular regions resulted in few baryonic droplets, producing weak IGM flow and galaxy flow. The result is the formation of galaxy group, such as the Local Group, which has fewer dwarf galaxies and lower density environment than cluster.

Clusters merged to generate tidal debris and turbulence, producing the baryonic droplets, the ICM (intra-cluster medium) flow, and the cluster flow. The ICM flow and the cluster flow directed toward the merger areas among clusters and particularly the rich clusters with high numbers of galaxies. The ICM flow is shown as the warm filaments outside of cluster [50]. The dominant structural elements in superclusters are single or multi-branching filaments [51]. The cluster flow is shown by the tendency of the major axes of clusters to point toward neighboring clusters [52]. Eventually, the observable expanding universe will consist of giant voids and superclusters surrounded by the dark matter halos.

In summary, the whole observable expanding universe is as one unit of emulsion with incompatibility between baryonic matter and dark matter. The five periods of baryonic structure development are the free baryonic matter, the baryonic droplet, the galaxy, cluster, and the supercluster periods as Fig. 13. The first-generation galaxies are elliptical, normal spiral, barred spiral, irregular, and dwarf spheroidal galaxies. The second-generation galaxies are giant ellipticals, cD, evolved S0, dwarf ellipticals, BCD, and TDG. The universe now is in the early part of the supercluster period.

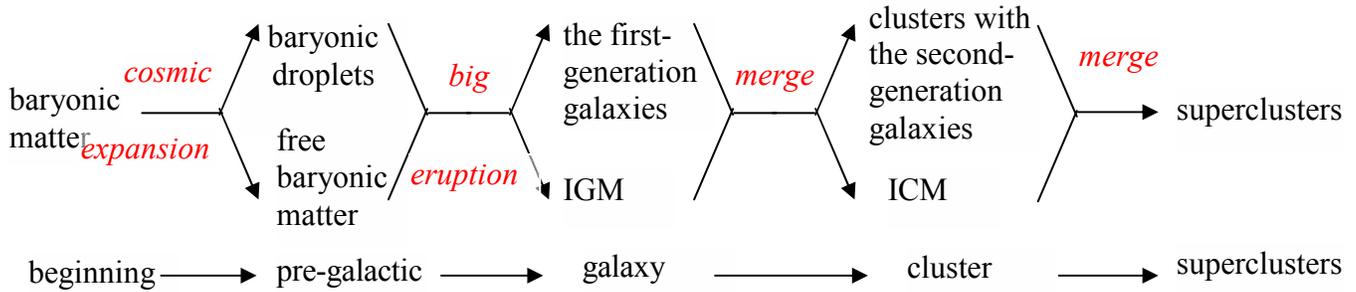

**Fig. 13**: the five levels of baryonic structure in the universe

### 4.3. Summary

The separation of dark matter and baryonic matter involves MOND (modified Newtonian dynamics). It is proposed that the MOND force is in the interface between the baryonic matter region and the dark matter region. In the interface, the same matter materials attract as the conventional attractive MOND force, and the different matter materials repulse as the repulsive MOND force between baryonic matter and dark matter. The source of the repulsive MOND force field is the incompatibility between baryonic matter and dark matter, like water and oil. The incompatibility does not allow the direct detection of dark matter. Typically, dark matter halo surrounds baryonic galaxy. The repulsive MOND force between baryonic matter and dark matter enhances the attractive MOND force of baryonic matter in the interface toward the center of gravity of baryonic matter. The enhancement of the low acceleration in the interface is by the acceleration constant, $a_0$, which defines the border of the interface and the factor of the enhancement. The enhancement of the low gravity in the interface is by the decrease of gravity with the distant rather than the square of distance as in the normal Newtonian gravity. The repulsive MOND force is the difference between the attractive MOND force and the non-existing interfacial Newtonian force. The repulsive MOND force field results in the separation and the repulsive force between baryonic matter and dark matter.

The repulsive MOND force field explains the evolution of the inhomogeneous baryonic structures in the universe. Both baryonic matter and dark matter are compatible with cosmic radiation, so in the early universe, the incompatibility between baryonic matter and dark matter increases with decreasing cosmic radiation and the increasing age of the universe until reaching the maximum incompatibility. The repulsive MOND force field with the increasing incompatibility results in the growth of the baryonic matter droplets. The three periods for the baryonic structure development in the early universe are the free baryonic matter, the baryonic droplet, and the galaxy. The transition to the baryonic droplet generates density perturbation in the CMB. In the galaxy period, the first-generation galaxies include elliptical, normal spiral, barred spiral, irregular, and dwarf spheroidal galaxies.

After reaching the maximum incompatibility, the growth of the baryonic droplets depends on the turbulence, resulting in the baryonic structure development of the cluster and the supercluster. In the cluster period, the second-generation galaxies include modified giant ellipticals, cD, evolved S0, dwarf elliptical, BCD, and tidal dwarf galaxies. The whole observable expanding universe behaves as one unit of emulsion with incompatibility between baryonic matter and dark matter through the repulsive MOND force field.

# 5. Summary

In this paper, the evolution of galaxies is by the incompatibility between dark matter and baryonic matter. Due to the structural difference, baryonic matter and dark matter are incompatible to each other as oil droplet and water in emulsion. In the interfacial zone between dark matter and baryonic matter, this incompatibility generates the modification of Newtonian dynamics to keep dark matter and baryonic matter apart. The five periods of baryonic structure development in the order of increasing incompatibility are the free baryonic matter, the baryonic droplet, the galaxy, the cluster, and the supercluster periods. The transition to the baryonic droplet generates density perturbation in the CMB. In the galaxy period, the first-generation galaxies include elliptical, normal spiral, barred spiral, irregular, and dwarf spheroidal galaxies. In the cluster period, the second-generation galaxies include modified giant ellipticals, cD, evolved S0, dwarf elliptical, BCD, and tidal dwarf galaxies. The whole observable expanding universe behaves as one unit of emulsion with increasing incompatibility between dark matter and baryonic matter. The properties of dark matter and baryonic matter are based on cosmology derived from the two physical structures: the space structure and the object structure. Baryonic matter can be described by the periodic table of elementary particles.

# 6. Reference

Email address: dy_chung@yahoo.com